\documentclass[apj,useAMS,usenatbib]{emulateapj}
\usepackage{times}
\usepackage{apjfonts}
\usepackage{amssymb}
\usepackage{epsfig}
\usepackage{rotating}

\begin{document}

\title{Detection of Signals from Cosmic Reionization using Radio
Interferometric\\ Signal Processing.} 

\author{A. Datta\altaffilmark{1,2}, S. Bhatnagar\altaffilmark{2} and
C.L. Carilli\altaffilmark{2}, } 

\altaffiltext{1}{New Mexico Tech, Socorro, NM 87801, USA}
\altaffiltext{2}{National Radio Astronomy Observatory, Socorro, NM
  87801, USA}

\email{adatta@nrao.edu}

\begin{abstract}
Observations of the HI 21cm transition line promises to be an
important probe into the cosmic dark ages and epoch of
reionization. One of the challenges for the detection of this signal
is the accuracy of the foreground source removal. This paper
investigates the extragalactic point source contamination and how
accurately the bright sources ($\gtrsim 1$ ~Jy) should be removed in
order to reach the desired RMS noise and be able to detect the 21cm
transition line. Here, we consider position and flux errors in the
global sky-model for these bright sources as well as the frequency
independent residual calibration errors. The synthesized beam is the
only frequency dependent term included here. This work determines the
level of accuracy for the calibration and source removal schemes and
puts forward constraints for the design of the cosmic reionization
data reduction scheme for the upcoming low frequency arrays like
MWA,PAPER, etc. We show that in order to detect the reionization
signal the bright sources need to be removed from the data-sets with a
positional accuracy of $\sim 0.1$~arc-second. Our results also
demonstrate that the efficient foreground source removal strategies
can only tolerate a frequency independent antenna based mean residual
calibration error of $\lesssim 0.2 \%$ in amplitude or $\lesssim
0.2$~degree in phase, if they are constant over each days of
observations (6 hours). In future papers we will extend this analysis
to the power spectral domain and also include the frequency dependent
calibration errors and direction dependent errors (ionosphere, primary
beam, etc).  

\end{abstract}

\section{Introduction}
Cosmic reionization corresponds to the transition from a fully neutral
to a highly ionized intergalactic medium (IGM), driven by ultraviolet
(UV) radiation from the first stars and black holes. The transition is
a key milestone in cosmic structure formation, marking the formation
of the first luminous objects. Reionization represents the last
major epoch of cosmic evolution left to explore. Study of the IGM,
galaxies, and quasars present during that time is a primary science
driver for essentially all future large-area telescopes, at all
wavelengths.  
 
Recent observations of the Gunn-Peterson effect, i.e.,  Ly$\alpha$
absorption by the neutral IGM, toward the most distant quasars ($z
\sim 6$), and the large scale polarization of the CMB, corresponding
to Thompson scattering during reionization, have set the first
constraints on the reionization process.  These data suggest
significant variance in both space and time, starting perhaps as far
back as $z \sim 14$ \citep{komatsu08} and extending to $z \sim 6$
\citep{fan06b}. Current probes of the reionization are limited:
present WMAP-V data indicates the 5~$\sigma$ detection of the E-mode of
polarization which rules out any instantaneous reionization at $z \sim
6$ at 3.5~$\sigma$ level.  For the Gunn-Peterson effect, the IGM
becomes optically thick to Ly$\alpha$ absorption for a neutral
fraction as small as $\sim 10^{-3}$. It has been widely recognized
that mapping the red-shifted HI 21cm line has great potential for
direct studies of the neutral IGM during reionization
\citep{furlanetto06}.    

There are number of upcoming low-frequency arrays whose key science
goal is to detect the HI 21cm signal from the Epoch of Reionization
(EoR). This includes the Murchison Widefield Array [MWA]
\citep{lonsdale09}, Precision Array to Probe Epoch of Reionization 
[PAPER] \citep{backer07} and Low Frequency Array [LOFAR]
\citep{jelic08}. One of the major challenges for all of these upcoming
arrays will be the removal of the continuum foreground sources in
order to detect the signals from the EoR. In this paper we discuss how
the radio interferometric imaging techniques are going to affect the
foreground source modeling and subsequent removal from the data-set in
order to search for the EoR signal. Recently, there has been
substantial research on foreground source modeling
(\citet{dimatteo02}, \citet{jelic08}, \citet{rajat08}, etc) at these
low frequencies. Similar effort has also been made in exploring
different techniques to remove the foregrounds from the EoR data-set
by \citet{morales06a,morales06b}, \citet{gleser08},\citet{bowman08},
\citet{liu08}, \citet{panos09} and \citet{harker09}. They
primarily focus on the removal of the sources that are fainter than a
certain $S_{cut}$ ($\sim 1$~Jy) level. Most of these works do not
consider the foreground sources brighter than $S_{cut}$ and how
accurately they need to be removed from the data-set by some real-time
calibration or modeling technique \citep{mitchell08} in the
UV-domain. In this paper we deal with the bright point sources above
$S_{cut}$ and the limitations that will be caused due to imperfect
removal of such sources. One of the objectives of this paper is to
demonstrate the effect of the frequency dependent side-lobes of the
synthesized beam on the foreground source removal strategies. The
effect of other frequency dependent terms like primary beam,
ionosphere, etc will be addressed in future papers.

In section-2, we discuss briefly our choice of model for the EoR
signal, i.e. the largest expected Cosmic Stromgren Sphere (CSS) which
represents the only signature that can be detected in the image domain
by the upcoming radio-telescopes. Section-3 presents the detailed
array parameters that have been used in the simulations. In section-4,
we discuss the foreground source model that has been used in every
simulation performed for this paper. Section-5 outlines the simulation
methodology and a possible data reduction procedure that might be
followed while processing the raw-data from the upcoming low-frequency
radio telescopes in order to extract the EoR signal. This procedure
may not be exactly the same as what will be actually implemented for
these upcoming telescopes. In section-6, we discuss the propagation 
of different forms of error through the radio interferometric data
reduction procedure. We consider two frequency independent errors: i)
position error in the Global Sky Model (GSM) that is used to remove
the bright sources above the $S_{cut}$ level and ii) residual
calibration error. We also present the results showing how these
errors propagate to the final residual (foreground-removed) spectral
image-cube. Finally, in the last section we discuss the implications
of the results from our simulations and our recommendations for the
upcoming low-frequency arrays in order to detect the signal of cosmic
reionization.

\section{EoR Signal - Cosmic Stromgren Spheres}

First generation low frequency arrays like MWA, PAPER, LOFAR, etc are
meant to detect three potential signatures of EoR: i) 3D power
spectrum,  ii) rare and large Cosmic Stromgren Sphere (CSS) and iii)
HI 21cm forest \citep{carilli02}. All of the above signatures require
the standard low-frequency radio-interferometric calibration and
imaging like self calibration, frequency dependent calibration and
source removal.    

For simplicity, we restrict ourselves to the large and rare Cosmic
Stromgren Sphere which are formed around the luminous quasars at the
end of reionization. These CSS form the only potential EoR signature
that can be detected in the image domain by the upcoming radio
telescopes.  

These CSS are rare and are expected to have a brightness temperature
of 20 ($x_{HI}$) mK with a physical size of $R_{phys} \sim 4.5$~Mpc
physical size, where $x_{HI}$ is the neutral fraction of the Inter
Galactic Medium (IGM). The physical size of these rare CSS has been  
derived from the Ly-$\alpha$ spectra \citep{fan06a}. According to
\citet{furlanetto06}, we derive the angular size of the CSS ($\Delta
\theta$) from the co-moving radius of CSS ($R_{com}$) :  
\begin{equation}
R_{com}=(1+z)R_{phys} = 1.9 \left(\frac{\Delta \theta}{1'} \right)
\left(\frac{1+z}{10}\right)^{0.2} h^{-1} ~Mpc
\end{equation}
This gives the angular scale of the CSS to be $\sim 16$~arc-minutes
along with a line-width of $\Delta \nu$ given by \citet{furlanetto06}:
\begin{equation}
R_{com}=(1+z)R_{phys} = 1.7 \left(\frac{\Delta \nu}{0.1MHz}
\right)\left(\frac{1+z}{10}\right)^{0.5} \left(\frac{\Omega_m h^2}{0.15}\right)^{-0.5} ~Mpc
\end{equation}
The line-width of the CSS translates to 2.5 MHz. Thus the total
flux density of the CSS is about 0.24 $x_{HI}$ mJy which gives a
surface brightness of:
\begin{equation}
S_B=22 (x_{HI}) \left(\frac{\theta_{beam}}{4.5'}\right)^{2} \mu~Jy~ beam^{-1}
\end{equation}
where $4.5'$ is the size of the synthesized beam for an array with
maximum baseline 1.5 Km at 158 MHz ($z=8$).

For a late reionization model, \citet{wyithe05} predict that the
$15^o$ field-of-view and 16 MHz bandwidth of MWA, will include atleast
one of these large and rare HII regions ($> 4$ Mpc at $z \sim
8$). Moreover, there is also the possibility of finding smaller HII
regions ($R \gtrsim 2$ Mpc) and up to $\sim 100$ fossil HII regions
due to nonactive AGN within the same field-of-view, depending on the
duty cycle \citep{wyithe05}.

\section{Array Specifications and Synthesized Beam}

Table 1 outlines the basic array parameters that we have adopted. We
note that most of these parameters reflect actual specifications for
the upcoming MWA-512 array. Figure 1 shows the array layout for the
512 element array with maximum baseline of 1.5 Km. This might not be
the final array design or specification for MWA. 

\begin{deluxetable}{cc}
\tablecaption{Array Specifications \label{tab_detector} }
\tablehead{\colhead{Parameters} &  \colhead{Values}}
\tablecolumns{2}
\startdata
No. of Tiles                    & 512 \\ \hline
Central Frequency               & 158 MHz ($z \sim 8$) \\ \hline
Field of View                   & $\sim$ 15$^o$ at 158 MHz. ($\propto \lambda$)  \\ \hline
Synthesized beam                & $\sim$ 4.5' at 158 MHz. ($\propto \lambda$) \\ \hline
Effective Area per Tile         & $\sim$ 17 $m^2$ \\ \hline
Maximum Baseline                & $\sim$ 1.5 km. \\ \hline
Total Bandwidth                 & 32 MHz \\ \hline
$T_{sys}$                       & $\sim$ 250 K \\ \hline
Channel Width                   & $\sim$ 32 kHz \\ \hline
Thermal Noise                   & $\sim$ 15.4 $\mu$~Jy/beam\\
      ~ ~                       & ($5 x 10^3$ hours $\&$ $2.5$ MHz) \\
\enddata
\tablecomments{Array parameters have been influenced by the MWA
specifications as mentioned in \citet{mitchell08} and
\citet{bowman08}. Original MWA Field-of-view is $\sim 25^{o}$ at 150 MHz.} 
\end{deluxetable}

\begin{figure*}[t!]
\centering
\epsfig{file=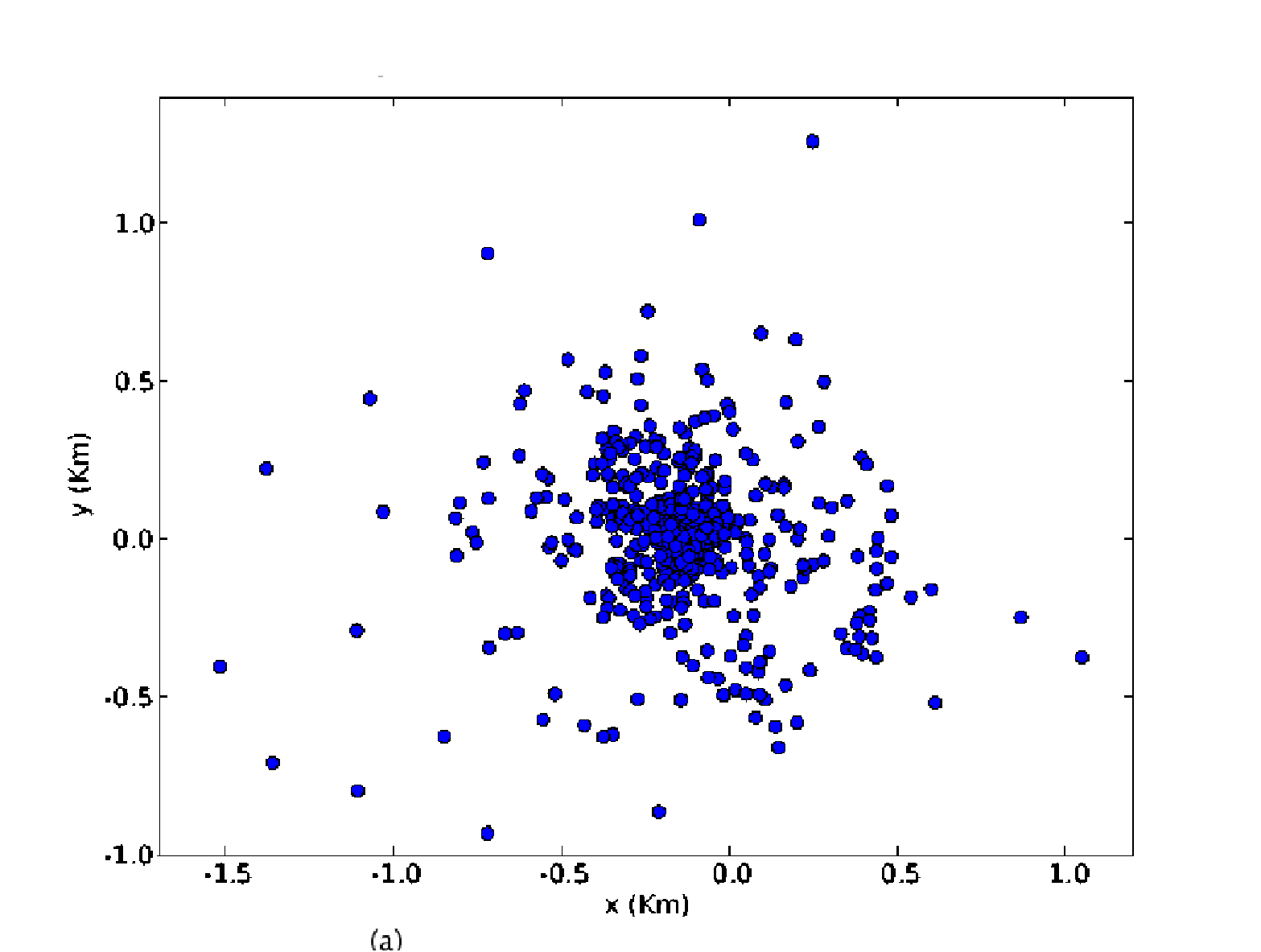,height=2.2truein}
\epsfig{file=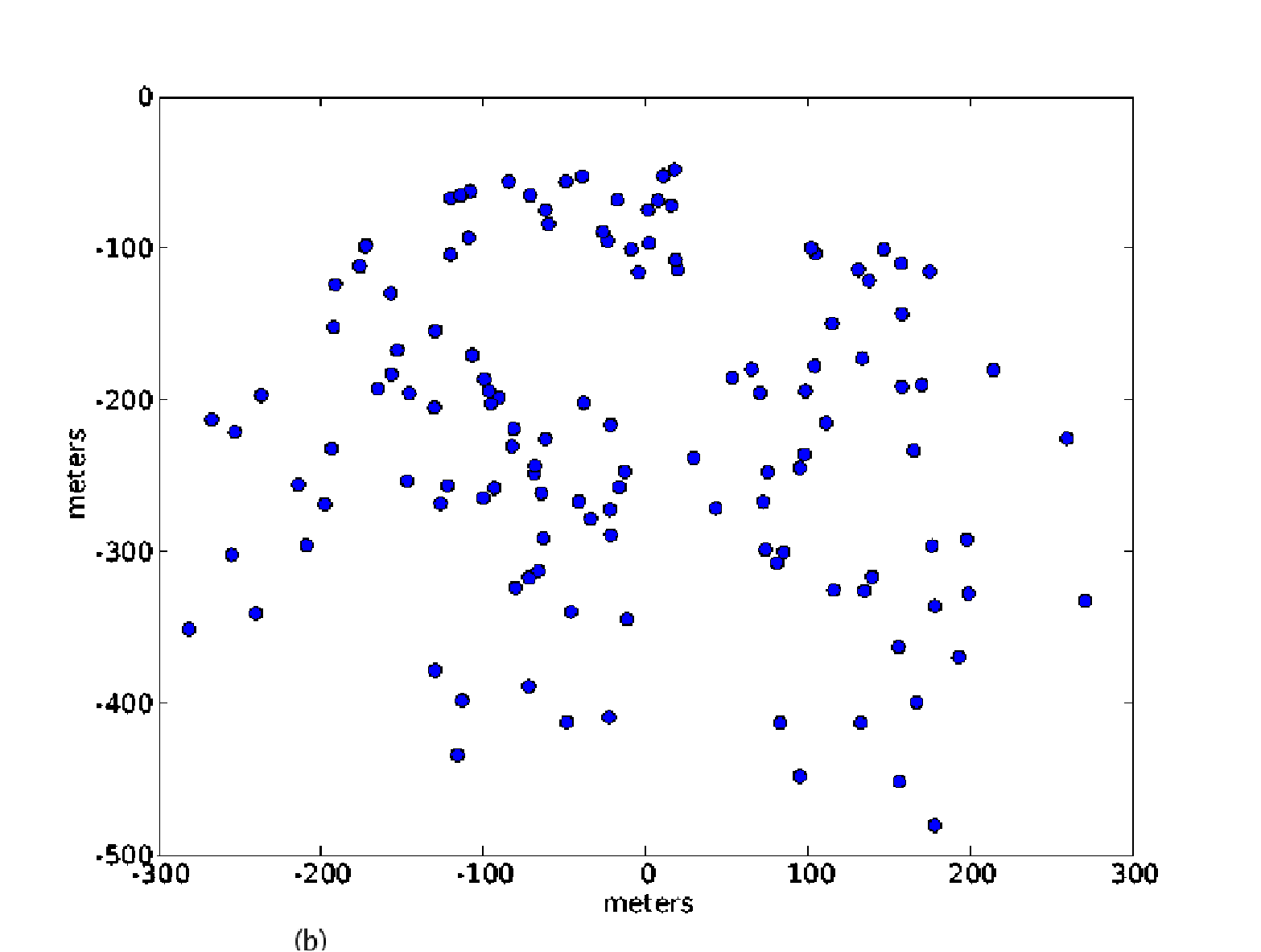,height=2.2truein}
\caption{{\bf Left:} Array layout for the 512 elements with maximum baseline
of 1.5 Km. {\bf Right:} Array layout for the 128 elements with maximum
baseline of 600m}
\label{fig:tszn}
\end{figure*}

In order to detect the signal from cosmic reionization, arrays like
MWA, PAPER, LOFAR, etc will have to overcome the thermal noise
limitation \citep{fan06b} :
\begin{equation}
\sigma_T = \left( \frac{1.9}{\sqrt{\Delta \nu_{kHz}~ t_{hr}}} \right)
\left(\frac{T_{sys}}{A_{eff} ~N_{ant}}\right) Jy~ beam^{-1}
\end{equation}
where $\sigma_T$ denotes the final RMS noise in the image from the
channel width of $\Delta \nu_{kHz}$ after $t_{hr}$ hours of
integration. $T_{sys}$, $A_{eff}$ and $N_{ant}$ denotes the system  
temperature, effective collecting area of each element/tile and number
of elements/tiles in the array. According to the above equation, the
thermal noise is 15.4 $\mu$ Jy/beam after $5 \times 10^3$ hours of
integration with $T_{sys} = 250$K \footnote { The value of
    $T_{sys}$ at these low frequencies is dominated by the sky
    temperature ($T_{sky} \propto \nu^{-2.6}$)
    \citep{furlanetto06}. We have adopted the above noise value for
    158 MHz. In practice, one should quote the noise figure for the
    lowest frequency edge which corresponds to the highest value for
    $T_{sky}$. } and channel width of 2.5 MHz, which is also the
spectral width of a CSS, as discussed in the previous section. 

Most of the upcoming low frequency telescopes will be
transit-instruments and will observe a field around its transit. Hence
we have used 6 hours of integrations for all the simulations, assuming
that the telescopes will observe a field between $\pm$~ 3 hours in
Hour Angle. Here, we have assumed that the field-of-view of the
instrument to be $15^o$. 

In this paper, we aim to demonstrate the case for the 512-element
array, as mentioned in Table 1. However, the simulations of
512-element are expensive. Hence, we have considered a simpler
geometry with 128-element array (figure 1), with a maximum baseline of
600 meters. This array design does not represent any of the
upcoming telescopes but has been adopted in order to simplify the
simulations. Apart from the effective area and synthesized beam value
the rest of the specifications for the 128-element array remains the
same as in Table 1. In our test simulations, we have used identical
residual calibration errors and created separate resultant spectral
image cubes for the two different array specifications (128-element
and 512-element). The RMS noise level in identical regions of these
two maps differed by a factor of $\sim 5$. The synthesized beam or the
PSF (point spread function) image from these two array configurations
shows that the RMS noise level in PSF side-lobes for the 512-element
is a factor of $\sim 5$ lower than that in the 128-element array. This
confirms that there is a scaling property between the 512-element and
the 128-element array. The same scaling property is also evident in
the results from our test simulations with the position errors. We
have exploited this property and have used the 128-element array for
all the simulations referred to in the later sections and have
subsequently scaled down the RMS noise values by the scaling factor,
in order to represent the same for the 512-element array. We also note
here that the thermal noise and the CSS surface brightness values that
are quoted throughout the paper corresponds to the 512-element array
configuration. Since we are dealing with point sources in our
foreground source model the change in the maximum baseline between the
two configurations will not affect our conclusions.        

\subsection{Synthesized Beam - PSF}

The only frequency dependent component in our simulations is the
side-lobes of the point spread function (PSF). In this section we
discuss the effects of different weighting schemes on the PSF shape
and its dependence on frequency.  

Figures 2 and 3 show the effect of three different weighting functions
on the PSF side-lobes for the 512-element array. These figures show
different plots from three well-known weighting schemes used in the
synthesis imaging: Natural, Uniform and Robust. \citet{dan99} showed
that the Robust weighting scheme smoothly varies between the Uniform
to Natural weighting schemes. Figure 3 shows how the near-by
side-lobes are almost coherent for two different channels separated by
32 MHz while the far-away side-lobes become more stochastic in
frequency. The effect of the different weighting scheme
shows similar results for the 128-elements as well but with a higher
PSF side-lobe level. The RMS value computed on the far away side-lobe
level of the PSF for the 128-element array is about $\sim 5$ times
higher than that in the 512-element array.

\begin{figure*}[t!]
\centering   
\epsfig{file=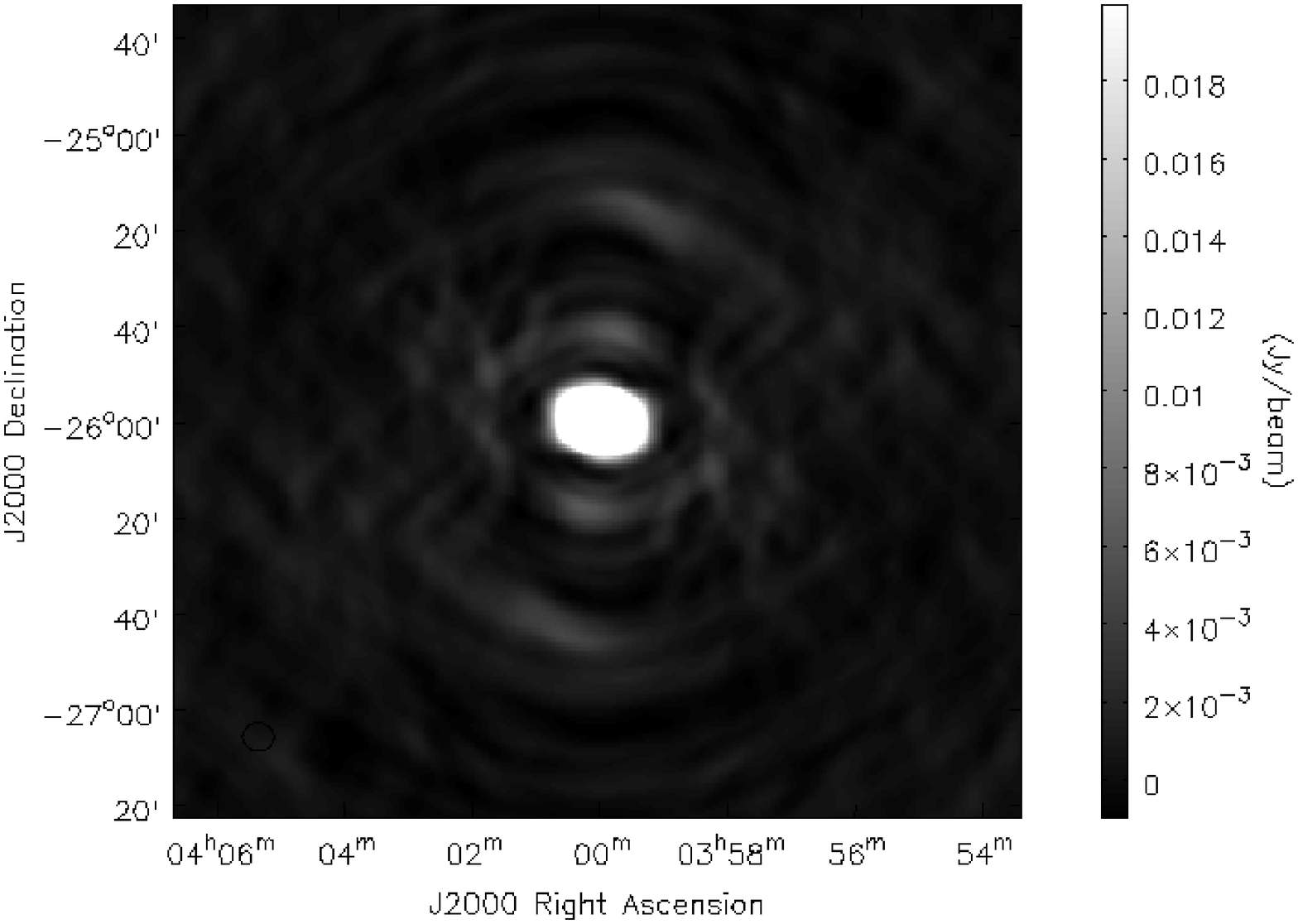,height=2.2truein}
\epsfig{file=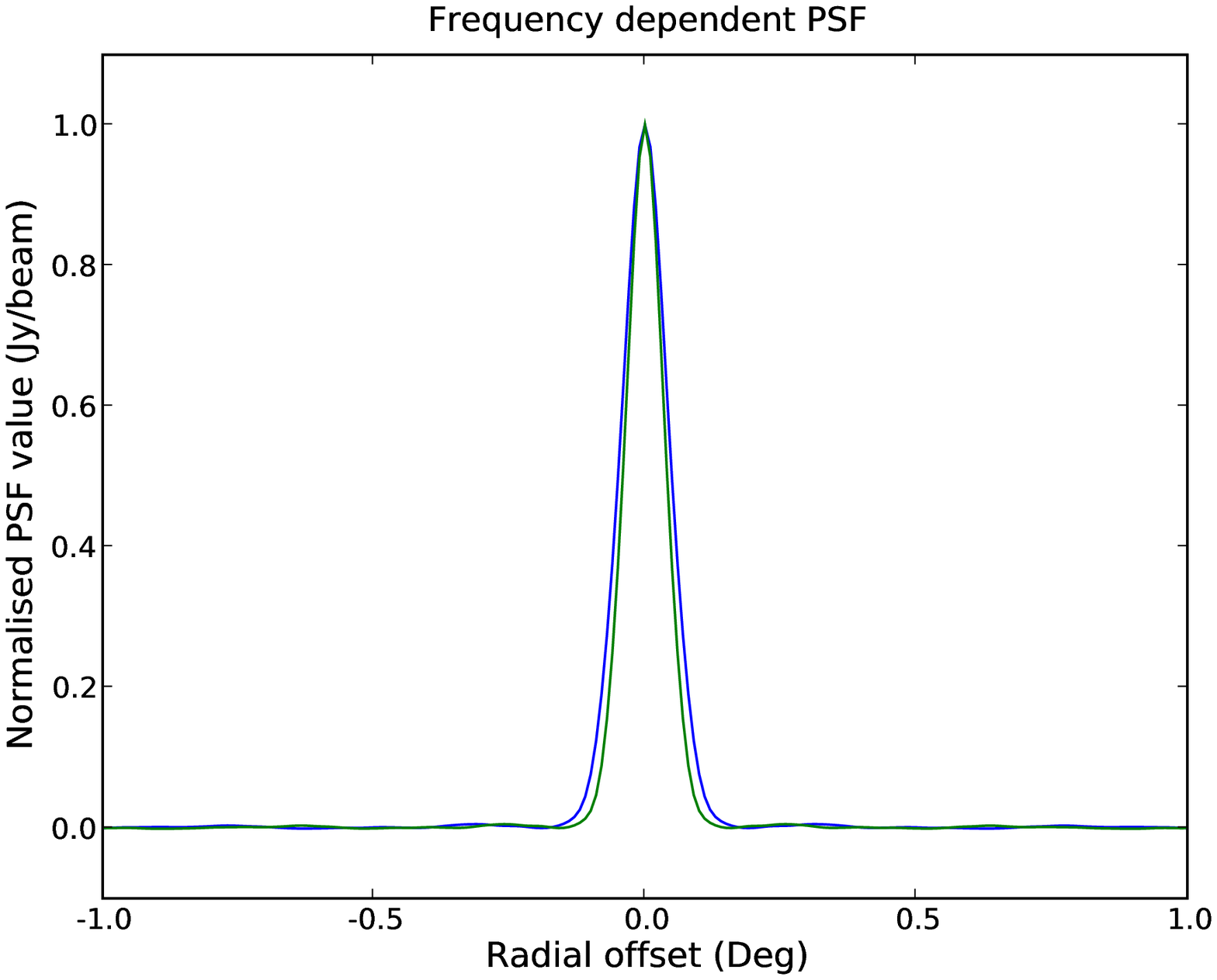,height=2.2truein}
\epsfig{file=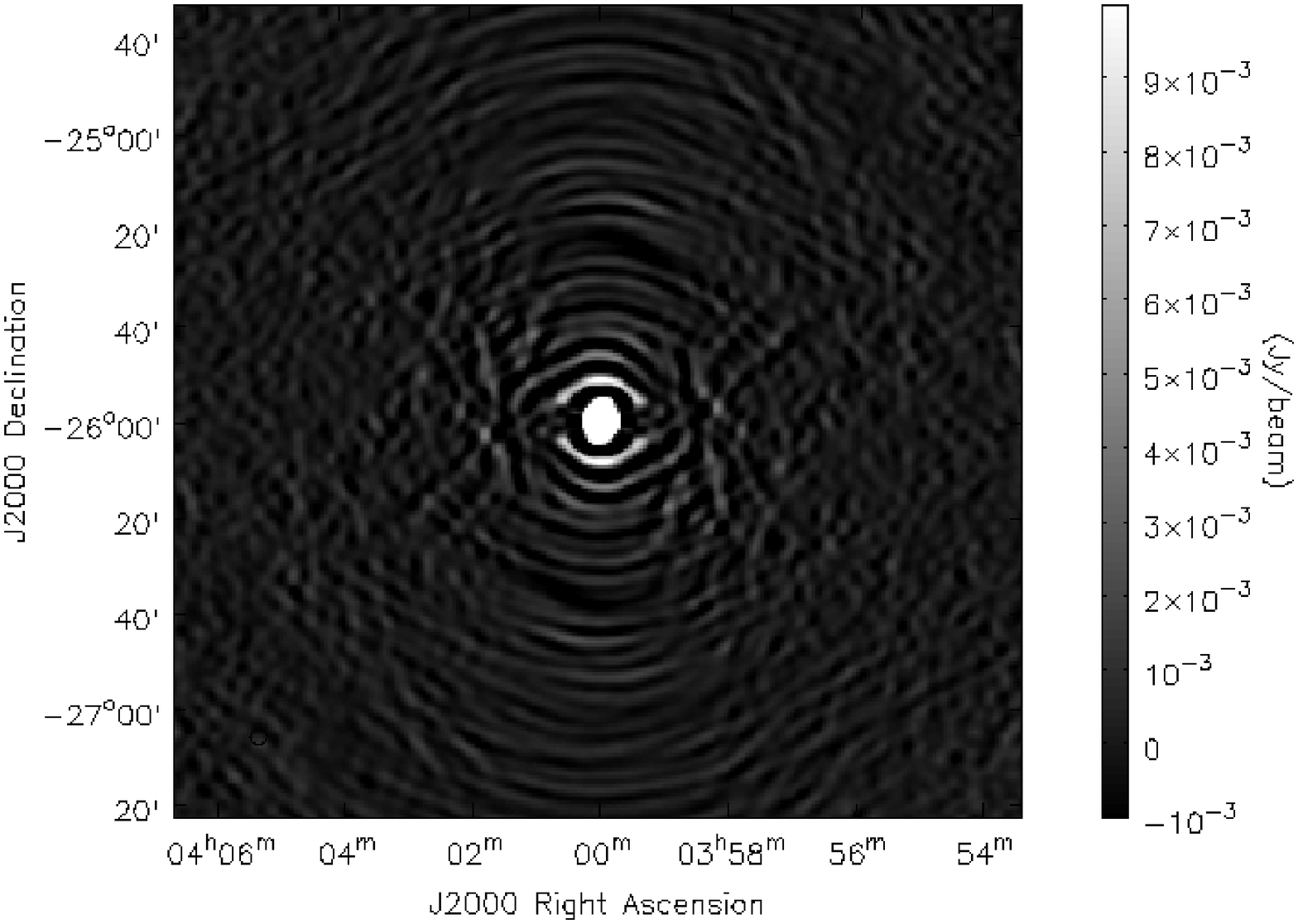,height=2.2truein}
\epsfig{file=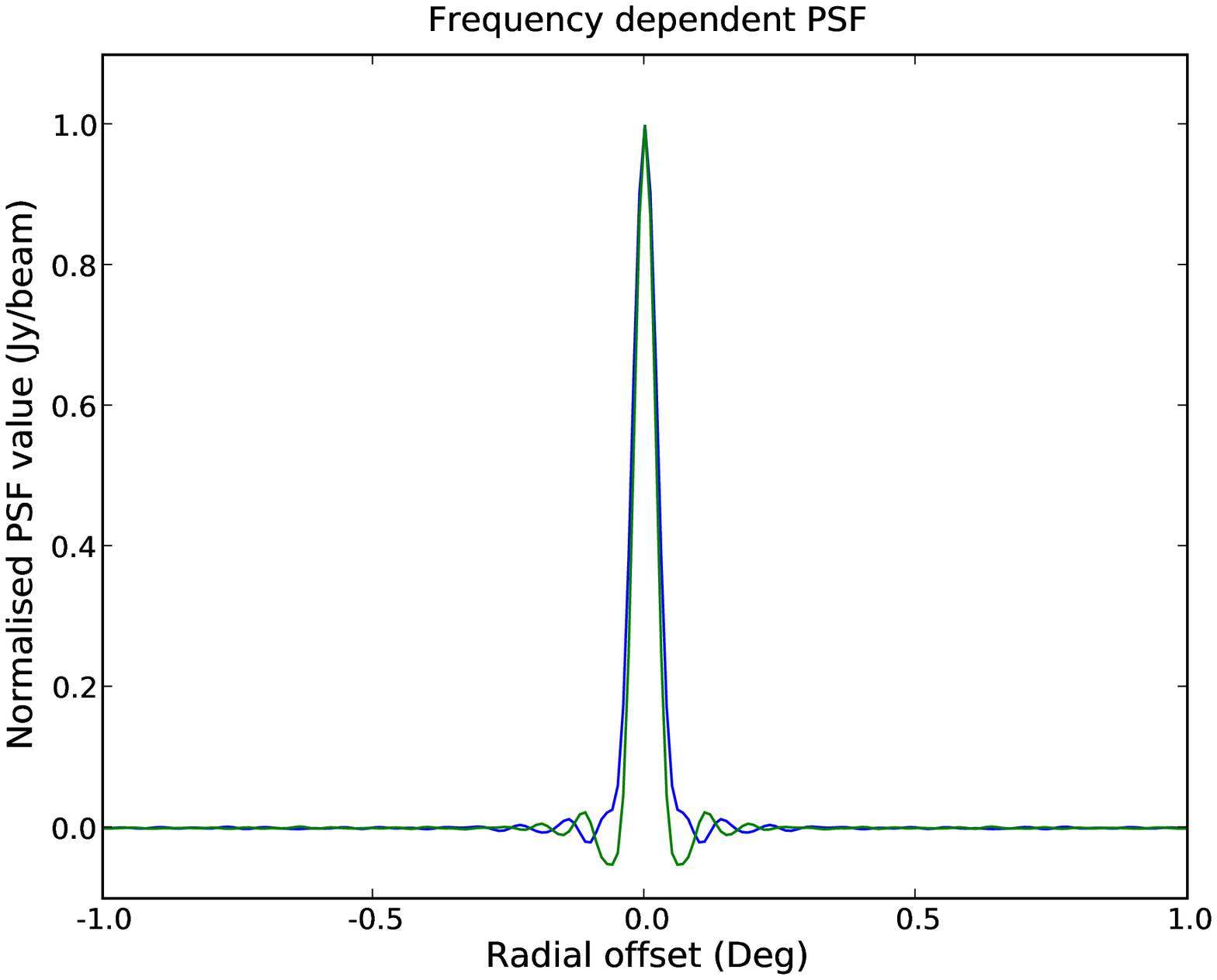,height=2.2truein}
\epsfig{file=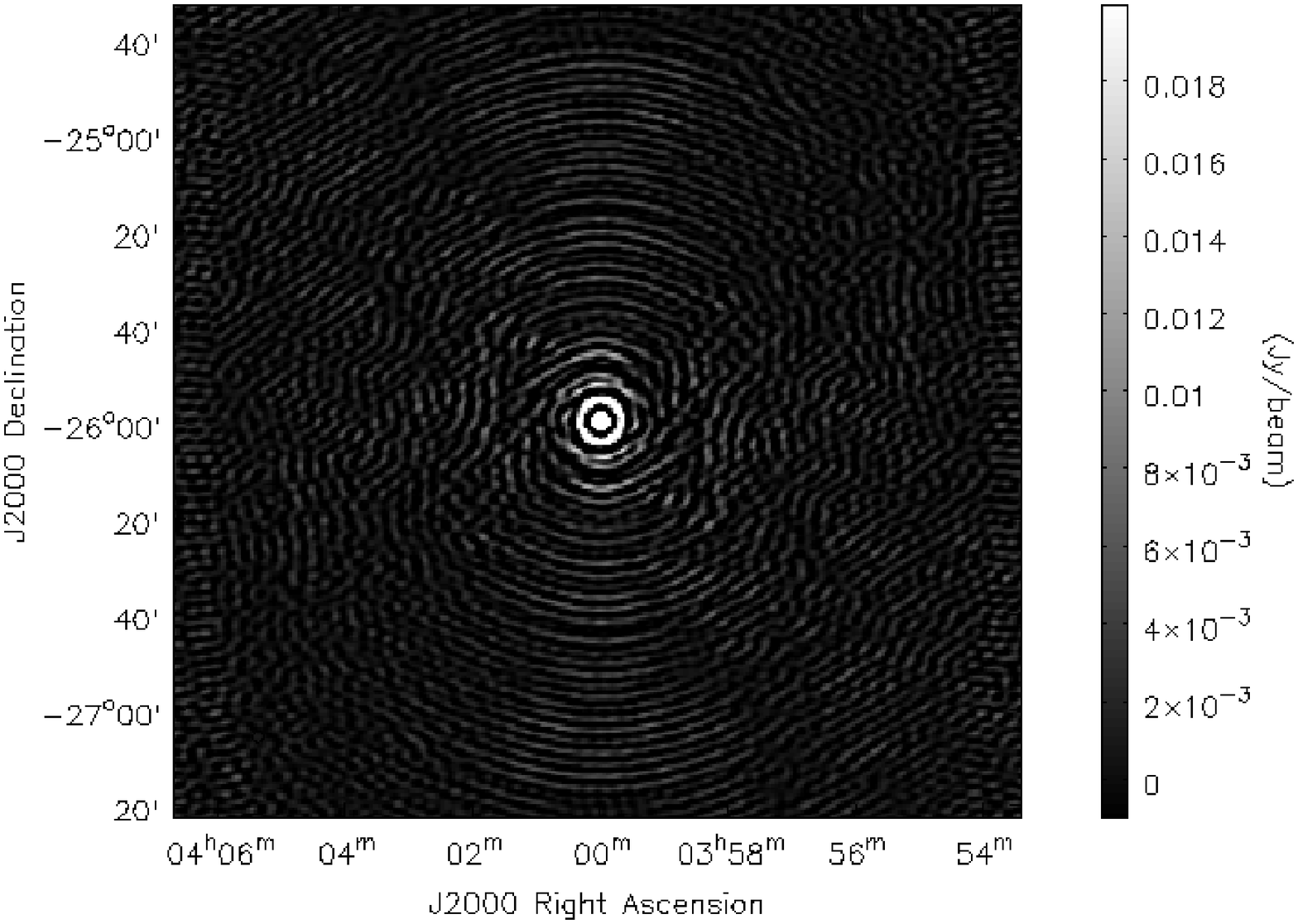,height=2.2truein}
\epsfig{file=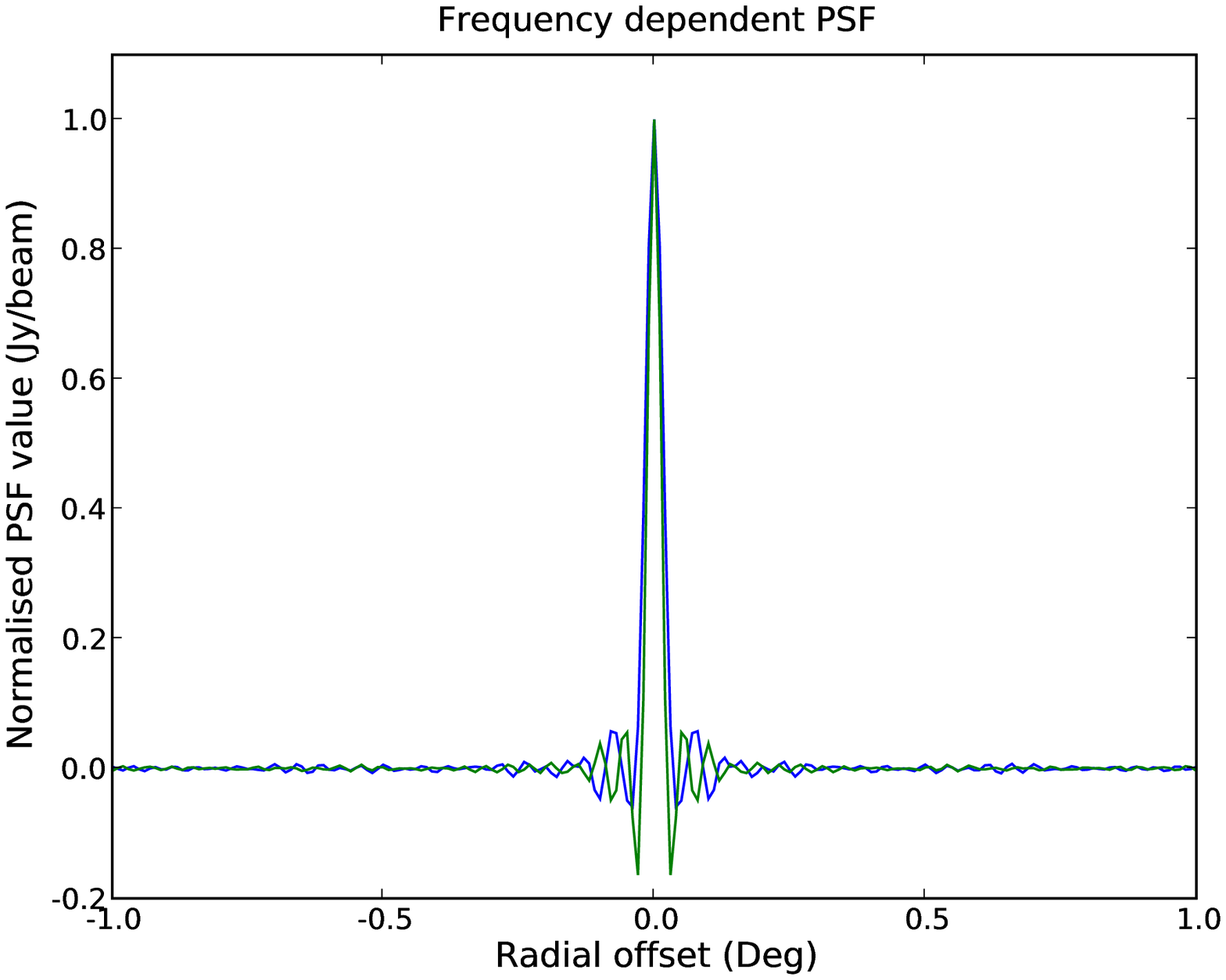,height=2.2truein}
\caption{Cuts through the Point Spread Function (PSF) for the
512-element array. Three different rows shows three different
weighting schemes. Natural (Top Row), Robust 0.0 (Middle Row) and
Uniform (Bottom Row). The left column shows the PSF image from the
different weighting schemes. The gray-scale levels shows the variation
in the intensity. The right column shows the cut through the inner
section of the PSF. The two different curves in all the plots shows the
PSF variation from two channels separated by 32 MHz in frequency.} 
\label{fig:tszn}
\end{figure*}

\begin{figure*}[t!]
\centering
\epsfig{file=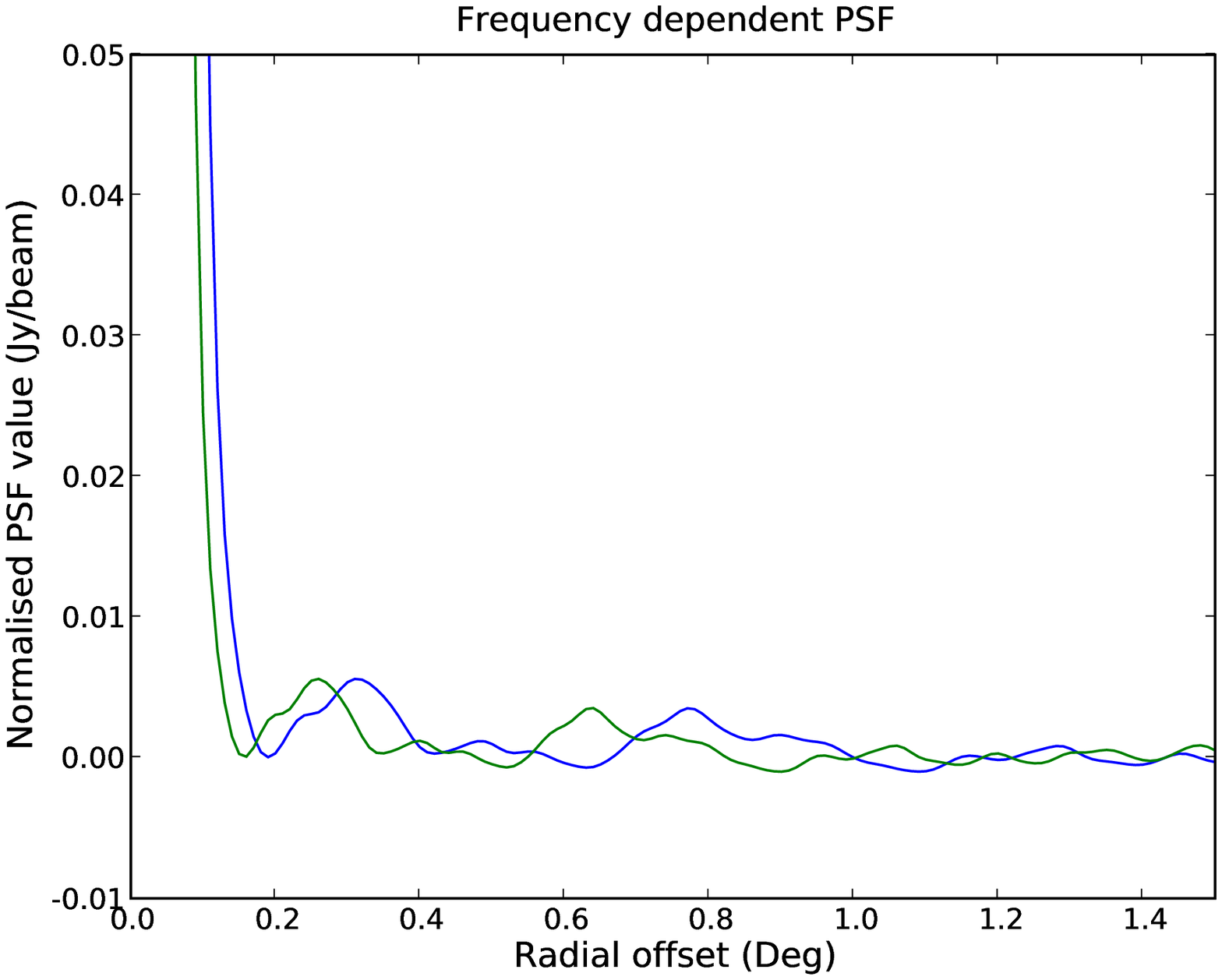,height=2.2truein}
\epsfig{file=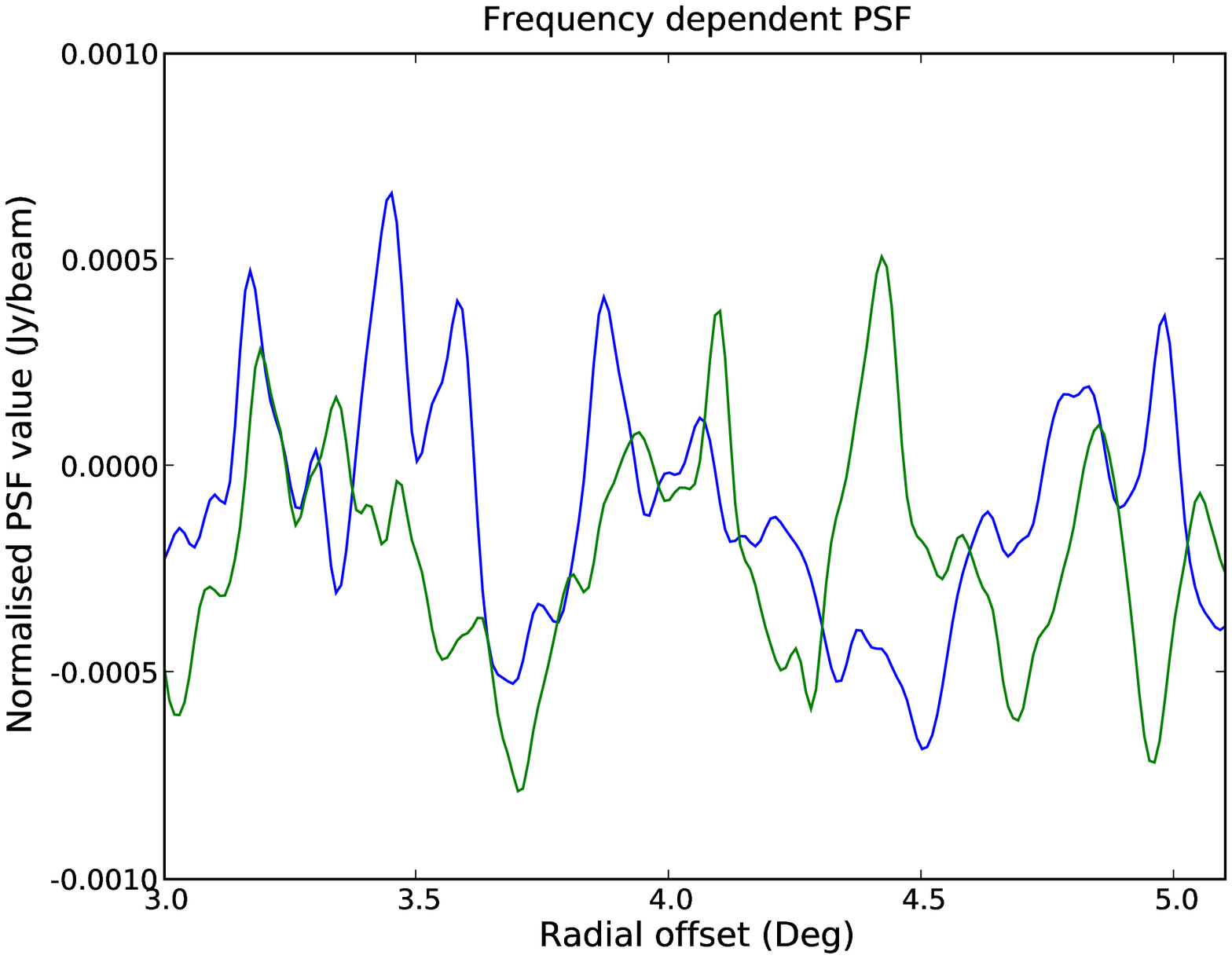,height=2.2truein}
\epsfig{file=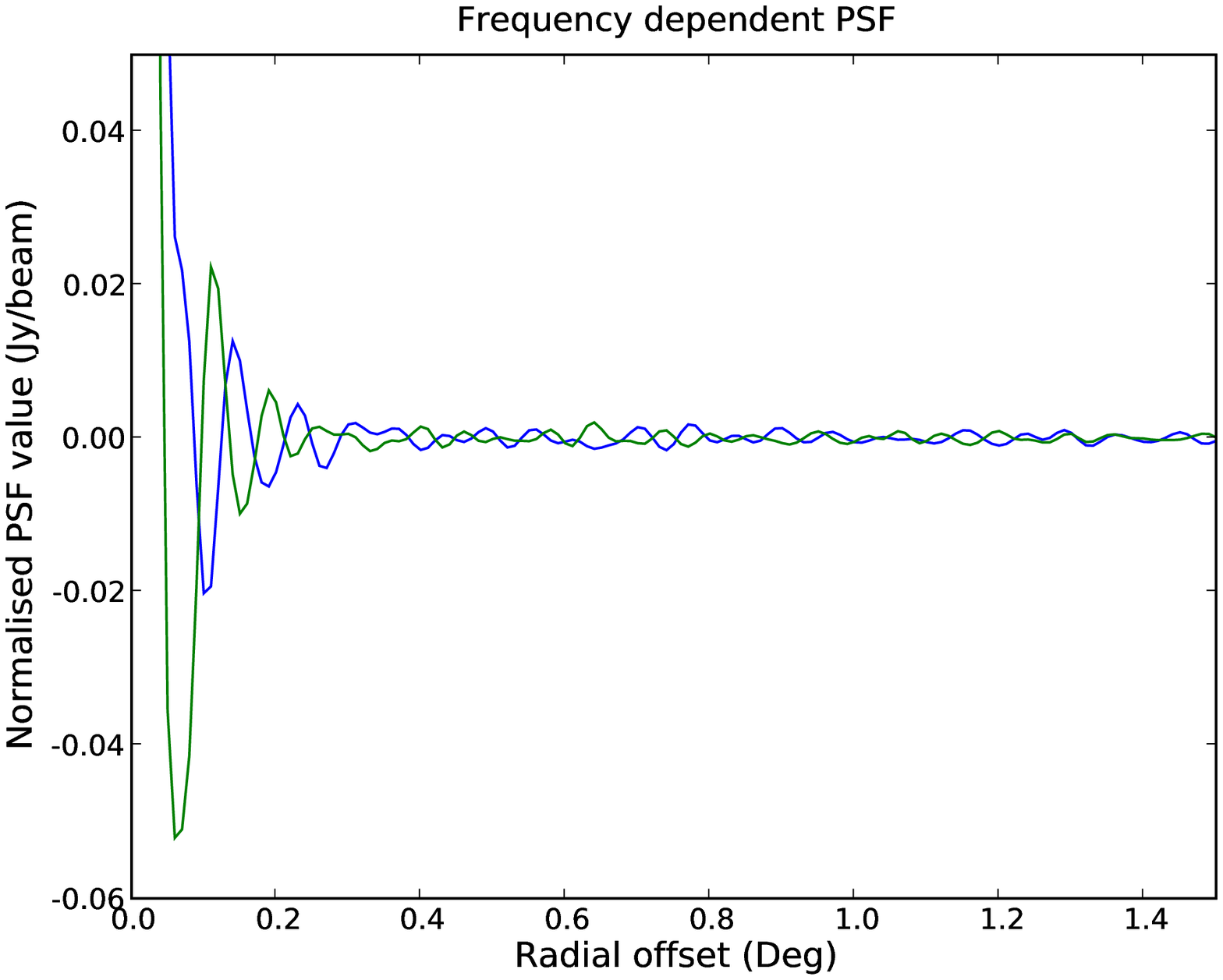,height=2.2truein}
\epsfig{file=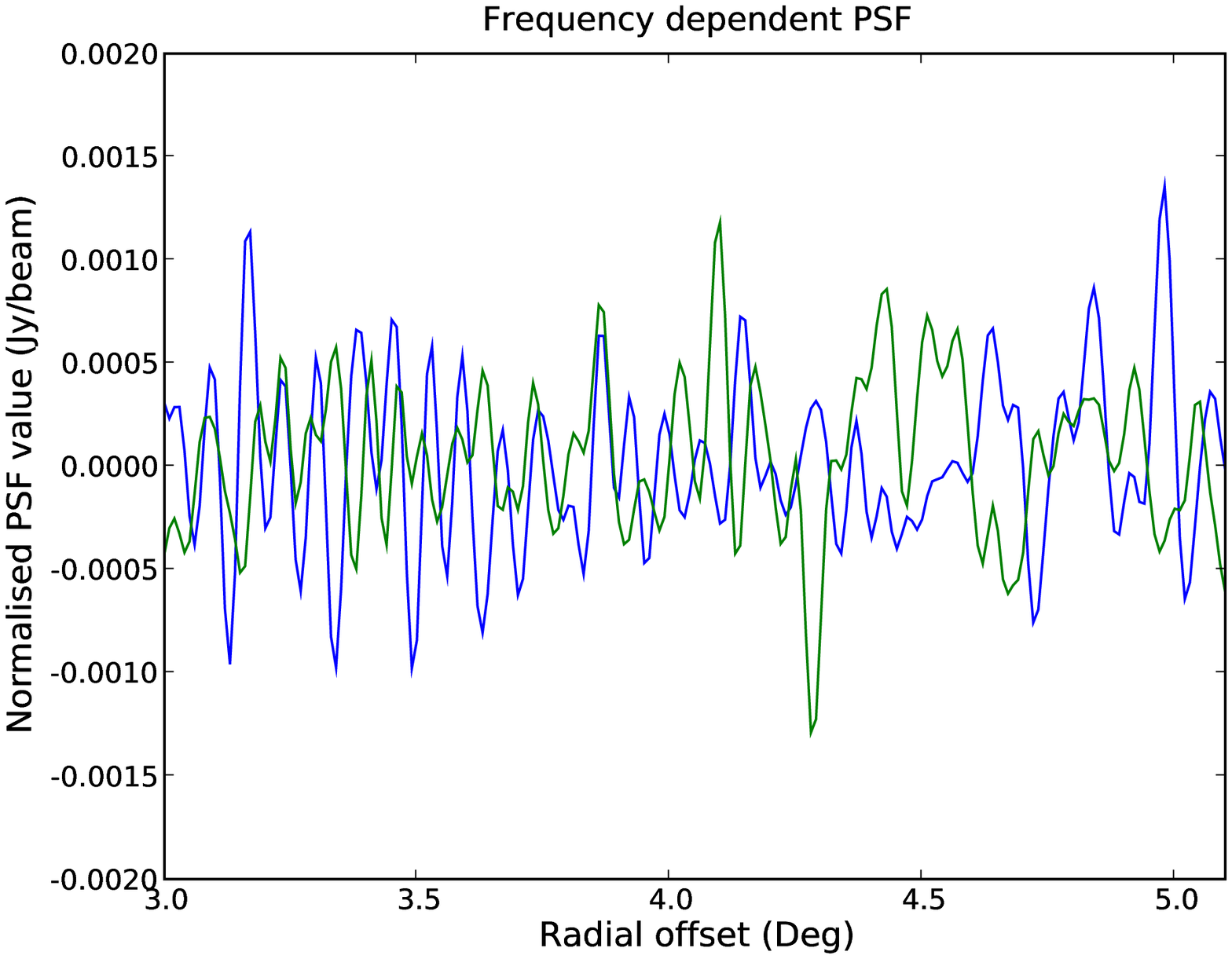,height=2.2truein}
\epsfig{file=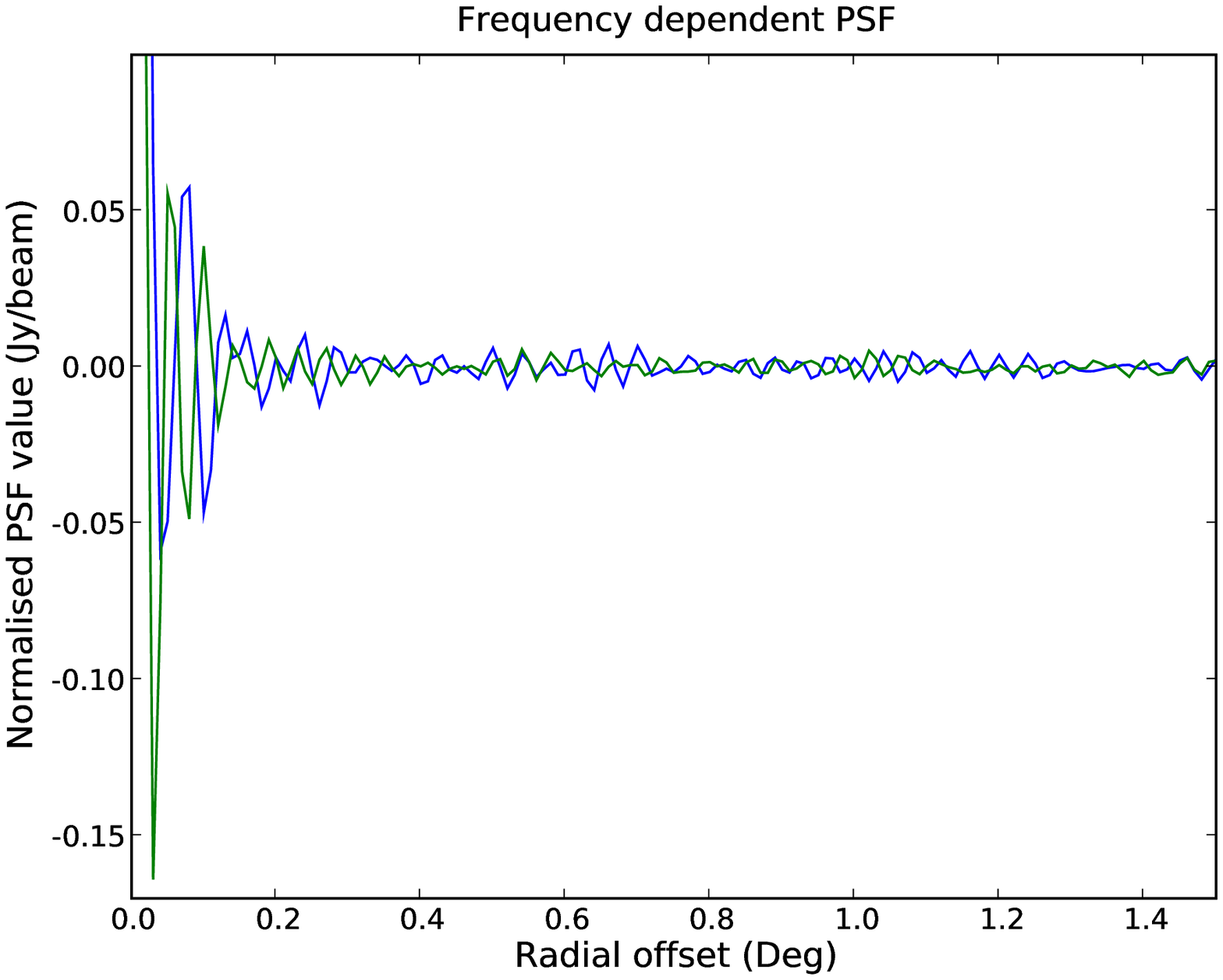,height=2.2truein}
\epsfig{file=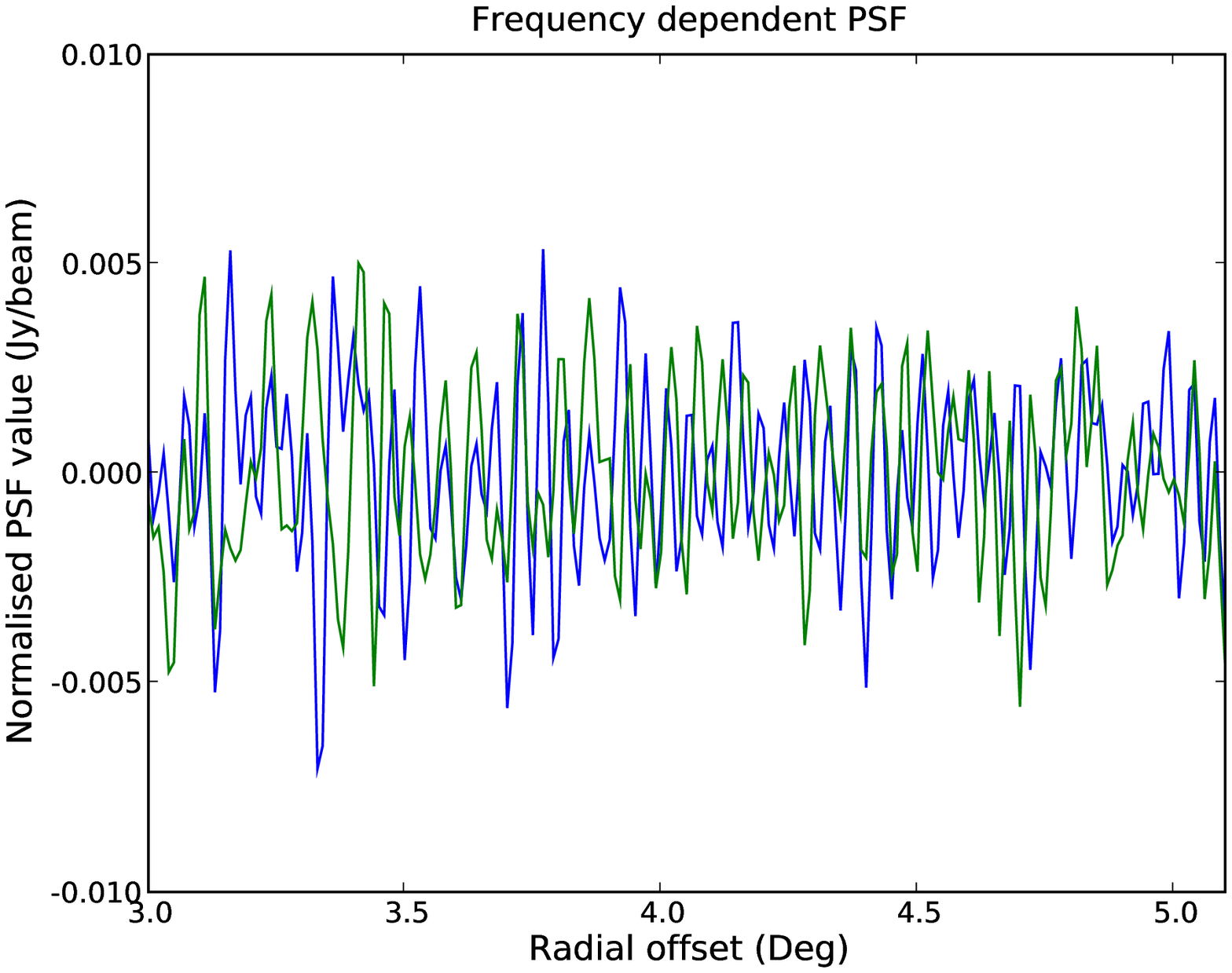,height=2.2truein}
\caption{Cuts through the Point Spread Function (PSF) for the
512-element array. Three different rows shows three different
weighting schemes. Natural (Top Row), Robust 0.0 (Middle Row) and
Uniform (Bottom Row). The left column zooms into the near side-lobe
levels while the right column shows the far-away side-lobe levels. The
abscissa shows the degree-offset from the center of the PSF. The two
different curves in all the plots shows the PSF variation from two
channels separated by 32 MHz in frequency.}
\label{fig:tszn}
\end{figure*}

We recognize that the choice of optimal weighting scheme depends on a
number of parameters based on the array configurations, telescope
hardware specifications and also on precise data reduction algorithms
used. Hence all the upcoming EoR experiments need an optimal
weighting scheme for their individual data reduction
procedures. However, exploring the optimal weighting scheme is beyond
the scope of this paper. For our work, we have adopted natural
weighting scheme. 

PSF side-lobe level reduces with increase in Hour Angle coverage. In
our simulations we have used six hours of observing time to produce
each data-set. However, in the actual experiment the observing time for
each data-set may be less than that. Hence the PSF side-lobe level
will also degrade accordingly. 

\section{Foreground Source Model - GSM}

Our foreground sky model only includes point sources. No extended
emission from galactic foreground is included as a part of the sky
model. Recently, there has been extensive research regarding the
foreground source removal dealing with sources below $S_{cut} = 1$~Jy
level. Most of these analysis implicitly assume that foreground 
sources above the $S_{cut}$ level are being removed perfectly. But in
reality any imperfect calibration will introduce artifacts in the
residual data after source removal. Hence in this paper our main aim
is to explore the level of accuracy needed in these calibration
procedures in order to ensure the residual errors from the strong 
foreground source removal do not obscure the detection of the signal 
from cosmic reionization. The choice of $S_{cut}$ level is not totally
arbitrary and will be discussed in the final section of this paper. 

The sky model is derived from the LogN-LogS distribution of sources
and is termed as the Global Sky Model (GSM) from now onwards. Since
our GSM only includes sources above 1 Jy, we follow the source count
from the 6C survey at MHz \citep{hales88}. The source count from the
6C survey is given as:  
\begin{equation}
N(>S_{Jy}) = 3600~ S_{Jy}^{-2.5} Jy^{-1} str^{-1} 
\end{equation}

For a field-of-view of $15^o$ the total number of sources ($> 1$~Jy)
equates to $\sim 170$, following the above power-law
distribution. The entire flux range between 1 - $10^3$~Jy was
  divided into several bins. The source population in each of these
  bins  has been predicted following the above LogN-LogS
  distribution. In order to assign fluxes to individual sources inside 
  a bin, a Gaussian random number generator has been used. The
strongest source in our GSM is $\sim 200$~Jy. The center of the field
is chosen such that it coincides with one of the cold spots in the
foreground galactic emission seen from the southern
hemisphere. Southern Sky is chosen since both the upcoming arrays MWA
and PAPER are being constructed in Western Australia. The exact field
center used for the GSM is 4 hours in Right Ascension and -26 degree
in Declination. In order to assign a position to each of these sources
within the field-of-view, another Gaussian random number generator has
been used which predicted the offset from the field center for
respective sources. In the GSM all the foreground sources are flat
spectrum, i.e. with zero spectral index ($\alpha = 0$).

Figure 4 shows the CLEANed image of the GSM used for all of the
simulations. The CLEAN algorithm used in this process is a wide-field
variant of the well-known Clark-CLEAN algorithm, which uses
w-projection algorithm for 3-dimensional imaging. 

In practice, the upcoming EoR telescopes should be using a similar
Global Sky Model (GSM) but created from the existing source catalogs
available for the part of the sky and frequency ranges of their
observations. These telescopes will use the GSM as the preliminary sky
model in order to detect and subsequently remove the bright foreground
sources in the observed data-set.

\begin{figure}[h!]
\epsscale{1}
\plotone{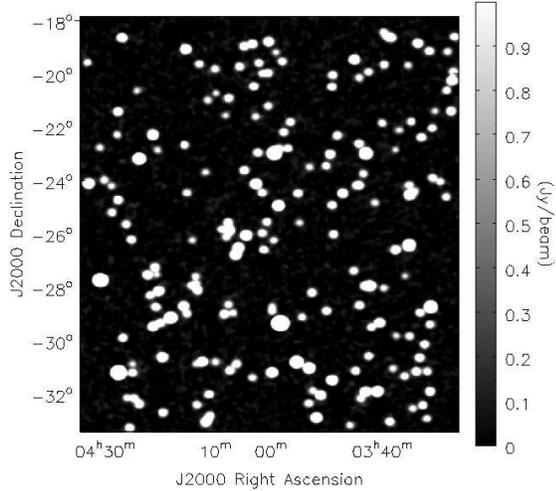}
\caption{Image of the Global Sky Model centered on RA=4h and
DEC=$-26^o$. Clark-CLEAN has applied to this image using w-projection
(256 planes) and Natural Weighting.} 
\end{figure}

\section{EoR Signal extraction - Data Reduction Procedure}

All the simulations have been performed within the CASA package
\footnote{http://casa.nrao.edu/}. As mentioned before, simulations for
both 512-elements as well as 128-elements have been conducted and
compared in the later sections. 

The $15^o$ field-of-view will include $\sim 170$ bright sources ($ >
1$~Jy). The individual flux densities of these foregrounds are $\sim
10^5 - 10^7 $ time higher than the signal from cosmic reionization
that these instruments are aiming to detect. So the challenge lies in
calibration and subsequent removal of such bright sources from the raw
data-sets. In order to add to the constraints, the data rates of most
of these telescopes (e.g. $\sim 19$~GB/s for MWA; \citet{mitchell08})
will not allow them to store the raw visibilities produced by the
correlator. Hence real-time calibration and imaging needs to be done
in order to reduce the data volume and store the final product in the
form of image cubes \citep{mitchell08}. The critical steps include
removal of the bright sources above the $S_{cut}$ level from the
data-sets in these iterative rounds of real-time calibration and
imaging procedure. As a result the residual image-cubes will not be
dominated by these bright sources and the rest of the foregrounds can
be removed in the image domain.     

However, the accuracy of the foreground source removal strategies are
strongly dependent on the data reduction procedure. The likely data
reduction procedure which will be followed by the upcoming telescopes
can be outlined as :

\begin{itemize}
\item The raw data-sets from the correlator will go through real-time
calibration and subsequent removal of the bright sources based on some
Global Sky Model (GSM), down to $S_{cut}$ level, in the UV domain. 
\item The residual data-sets will be imaged and stored as a cube for
the future processing and removal of sources which are below
$S_{cut}$. 
\end{itemize}

However, due to imperfect calibration during the process of real-time
calibration and source removal if any systematic error is incorporated
in the residual image-cubes they cannot be minimized by simply
averaging over a large number of such residual image-cubes. Hence to
understand the level of accuracy required in the calibration procedure
to reduce the residual systematic errors in the data-set is critical.

Here, we outline the reduction pathway that have been followed
throughout the paper :

\begin{itemize}
\item The observed visibilities ($V_{ij}^{Obs}$ or
$V_{ij}^{GSM_{perfect}}$) are simulated for 6 hours ($\pm 3$~hours in
Hour-angle) from the GSM (with no position or flux error) and the
array configuration from Table: 1.  

\item Model visibilities ($V_{ij}^{mod}$) are also generated using the
same GSM but they are now corrupted with either GSM position errors or
with residual calibration errors. 

\item Model visibilities are subtracted from the observed visibilities
to residual visibilities ($V_{ij}^{res}= V_{ij}^{obs}- V_{ij}^{mod}$).
This step will be referred to as UVSUB \citep{cornwell92}.

\item We fit a third order polynomial in frequency to the dirty
image-cube ($I^{res}$) formed from $V_{ij}^{res}$. This step will be
referred to as IMLIN \citep{cornwell92}. 

\end{itemize}

After the IMLIN step, we investigate the RMS noise in the residual
image cube. We have tried to use different orders of polynomial for
the IMLIN step but our conclusions did not change significantly. This
may be mainly due to the fact that the only frequency dependent term
in these simulations is the PSF. However, it should be mentioned in
this context that on using higher order polynomials in the IMLIN step
we will be taking out relevant structures at those scales in the
frequency space \citep{mcquinn06}. Since the reionization signal is a
spectral signature with a width $\lesssim 2.5$~MHz (size of the
largest possible CSS), using a very higher order polynomial in the
IMLIN might remove the signature itself.     

We have also explored UVLIN \citep{cornwell92}, which fits a
polynomial in the UV domain to $V_{ij}^{res}$. But UVLIN works
perfectly only within a small field-of-view depending on the channel
width in frequency \citep{cornwell92}. Hence, in our proposed
reduction pathway we are using UVSUB followed by IMLIN only.

\section{Error Propagation}

In this section we describe the effects of different errors arising
from radio interferometric data processing and how they propagate
and show up as artifacts in the final residual image. Here, we
consider effects of the errors arising from the imperfect GSM and
residual calibration errors on removal of strong foregrounds.

\subsection{Error due to approximate GSM - Simulations}

The source positions in the Global Sky Model (GSM) are accurate up to
a level, e.g. the NVSS catalog has an accuracy of $\sim 6$~arc-seconds
in positions of the sources detected \citep{nvss02}. The real-time
calibration procedure for MWA like telescopes might implement some
method to iteratively solve for the positions of bright sources taking
the initial value from the GSM, e.g. Asp-CLEAN \citep{sanb04}. An
iterative scheme will also converge with some residual position and
flux errors in the model. Hence, any error in the final GSM position
or flux will contribute to imperfect source removal. The resultant
artifact will limit the dynamic range \footnote{Dynamic range of an
  image is defined as the ratio of the peak brightness on the
  image and the RMS noise in a region of the image with no sources.}
in the residual data-set and obscure the detection of the faint signal
from reionization. The main aim of the simulations in this section is
to determine the accuracy of the GSM (position and flux) required in
order to reach the required RMS noise level. The GSM position errors
considered in the following simulations are actually systematic
residual errors after the iterative position determination. Here, we
deal with residual GSM errors in position and will also mention
briefly the effect of the GSM flux errors in the residual data-set.

In order to introduce the position error in the GSM we have added an
error term in the Right Ascension ($\alpha$) of each of the sources in
the GSM. This error term is derived from a Gaussian distribution:-
\begin{equation}
\epsilon(\alpha) = \frac{1}{\sigma \sqrt{2 \pi}} e^{\frac{\alpha^2}{2
\sigma^2}}
\end{equation}
where $\sigma$ denotes the error level in the position, i.e. for NVSS
catalog $\sigma = 6$~arc-seconds. 

The steps in the simulation performed in order to explore the effect
of the GSM error in position is as follows:

\begin{enumerate}
\item $V_{ij}^{Obs}$ are simulated.

\item Predict model visibilities from the GSM model with
  source-position error ($V_{ij}^{GSM_{imperfect}}$).  

\item Subtract the model visibilities from the observed visibilities
  to get $V_{ij}^{Res}= V_{ij}^{GSM_{perfect}} -
V_{ij}^{GSM_{imperfect}}= V_{ij}^{GSM_{error}}$. 

\item Make $I^{res}$ and apply IMLIN as mentioned befor.e

\end{enumerate}

In order to explain the above mentioned steps in detail we refer to
figure 5, which demonstrates the procedure of the reduction followed
in the simulations involving GSM errors. Figure 5a shows the dirty map
(peak value: $\sim 2 \times 10^{-3}$~Jy/beam) from the residual
visibilities where the model visibilities where corrupted with a GSM
position error of 0.01 arc-second. Figure 5b (peak value: $\sim 8
\times 10^{-5} $~Jy/beam) is generated after applying IMLIN to the
residual image. The gray-scale color-bars represent the variation of
the intensity in two images.   

\subsection{Position Error in the GSM - Results}

Here we discuss the implications of the results from the simulations
with the GSM errors. Figure 6a represents the variation in the RMS
noise level of three regions in : [1] True Image of
$V_{ij}^{Obs}$ (figure 4), [2] Image of $V_{ij}^{Res}$ after UVSUB
(figure 5a) and [3] Image after IMLIN (figure 5b). The GSM position
errors used in this case is 0.01 arc-second. 

In figure 6b, the RMS noise value in three different parts of the
field-of-view have been plotted as a function of different residual
GSM position errors. The RMS values quoted here are from the images of
residual visibilities after UVSUB [2]. Hence the solid line denotes
the trend in the decrease of the RMS noise in the UVSUB image 
with the decrease in the GSM position error. The magnitude of the
position error (value of $\sigma$ as in equation 6) has been varied
from 6 arc-seconds (as in NVSS catalog) down to an extremely low value
of $10^{-4}$ ~arc-seconds. The original simulations were performed with
128-element array and the resulting RMS values are scaled down by
factor of 5 in order to represent the same for the 512-element
array. This scaling property has already been discussed in details in
section-3 of this paper. According to the figure 6b it is evident that
an accuracy of 0.1 arc-second in GSM position is needed to achieve the
required RMS noise in order to detect the reionization signal after
removal of strong foreground point sources. The solid curve in figure
6b represents the mean curve of the RMS noise level variation and has
a mean slope of $\sim 1$ in the log-log space. This means that a
decrease in the RMS noise level by another order of magnitude is
possible on achieving an extra order of magnitude accuracy in the GSM
position estimates.    

\citet{maxim04} have shown similar results in demonstrating the effect
of the pixelation error on the dynamic range of the image. Recently,
\citet{cotton08} have shown that a image made from a data-set using
the VLA (27-element) obtains a dynamic range limitation of $10^5$ when
the sources are shifted by 0.01 pixel in the image plane. This result
is similar to the result quoted in figure 6b, where it is evident that
GSM position error of 1 arc-second ($\sim 0.01$ pixel) results in an
RMS noise of $\sim 10^{-4}$ Jy/beam corresponding to a dynamic range
(DR) of $\sim 10^6$. The better PSF side-lobe level in 512-element
array over VLA27-element array causes an extra order of magnitude gain
in the dynamic range in the case of the former. 

Here, we have considered only systematic residual GSM position errors
which are the same from one day to the another. This will not reduce
after averaging over residual image-cubes from successive days of
observations. The real-time calibration system might implement an
iterative scheme (e.g. Asp-CLEAN, \citealp{sanb04}) to determine the
source positions. However, unless the systematic error determined by
any such scheme is smaller than 0.1 arcsecond, averaging over long
lengths of time will not mitigate the errors due to imperfect GSM. The
GSM position error limit that we report in this paper is thus also
applicable for the specifications of the accuracy that any such scheme
delivers.  

\begin{figure*}[t!]
\centering
\epsfig{file=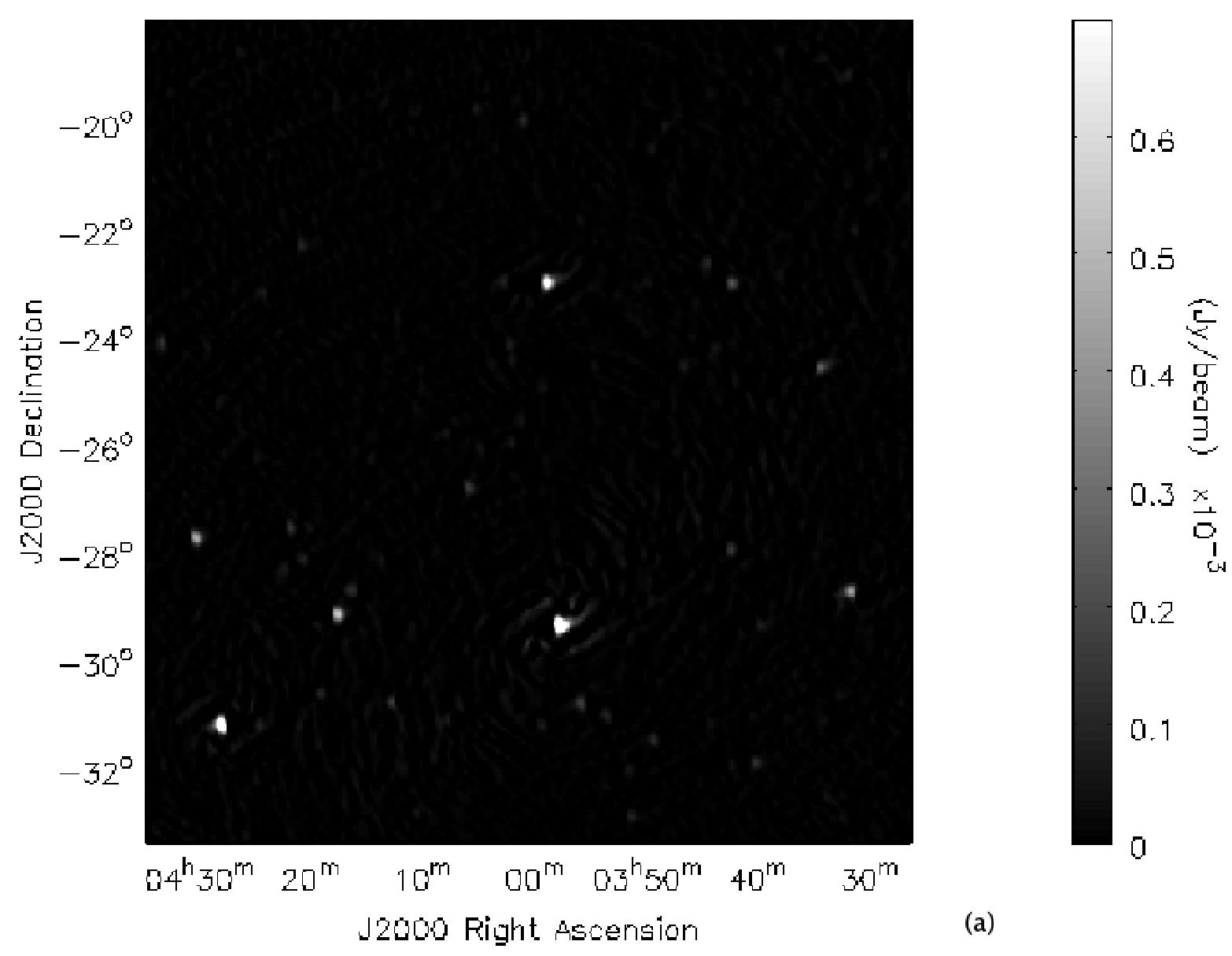,height=2.6truein}
\epsfig{file=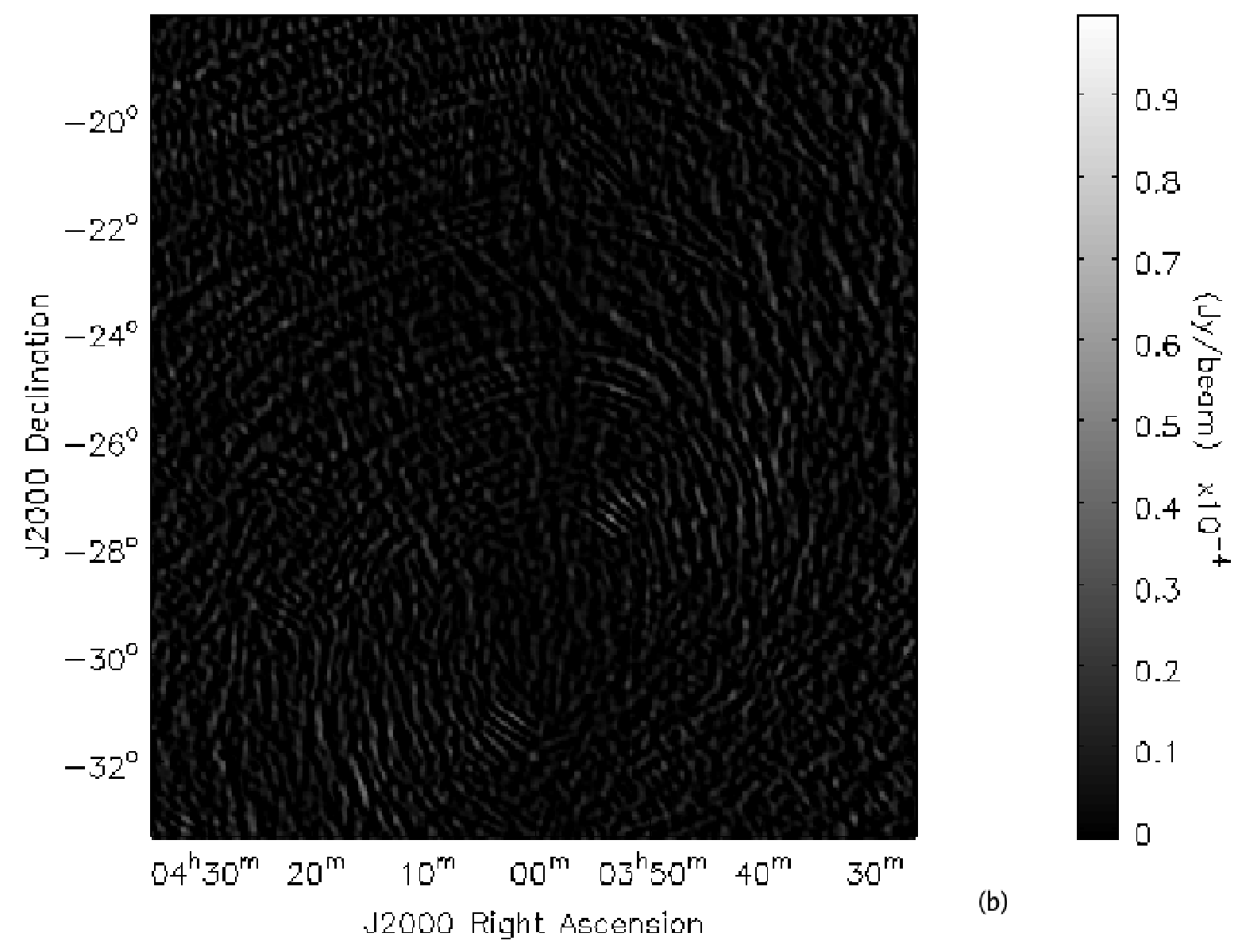,height=2.6truein}
\caption{{\bf (a)}Dirty UVSUBed image of the field made from
  $V_{ij}^{res}$ after the imperfect GSM ($V_{ij}^{GSM_{imperfect}}$)
  has been subtracted from the perfect data-set ($V_{ij}^{obs}$). {\bf
  (b)} The image of the field after the IMLIN has been applied to the
  UVSUBed image in figure a.}  
\end{figure*}

\subsection{Flux Error in the GSM - Results}

In this section we discuss the results of introducing GSM flux error
in our simulations. GSM flux errors have been derived from a Gaussian
random  distribution (similar to equation 6) with mean zero and
$\sigma \sim 0.1 \%$ of flux values of respective sources. Similar to
the steps of simulations discussed in section 6.1, the GSM model with
flux error is used to corrupt model visibilities. The residual image
cube shows sources with flux values of $\lesssim 0.1 \%$ of the
original sources in the same position. Since the GSM contains only
sources between 1 Jy and $\sim 200$~Jy, the maximum residual flux
density in the residual image is $\lesssim 0.2$ Jy/beam and can be
denoted as sources below the $S_{cut}$ limit \citep{mitchell08}. The
removal of such sources are beyond the scope of this paper.   

\begin{figure*}
\epsfig{file=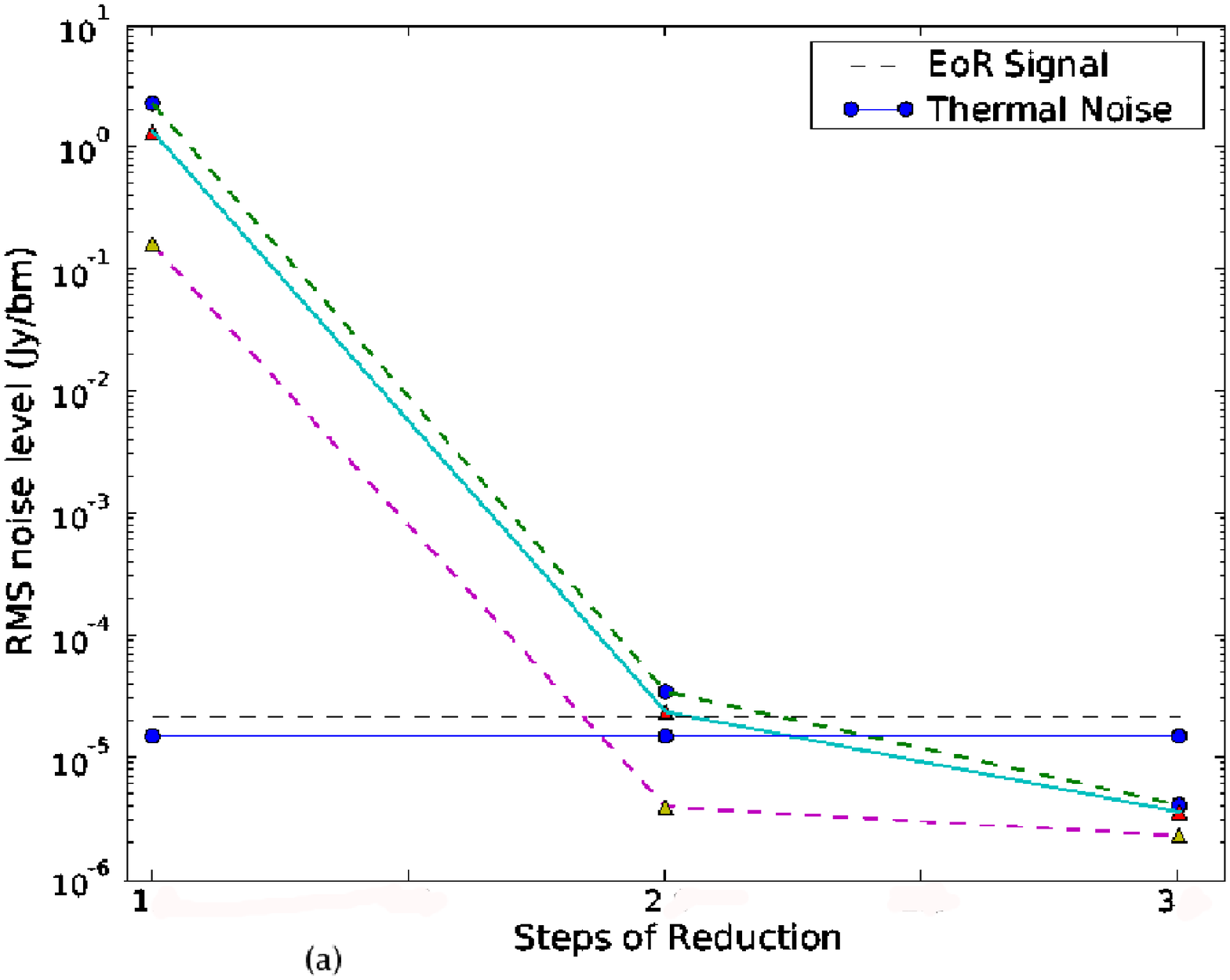,height=2.6truein}
\epsfig{file=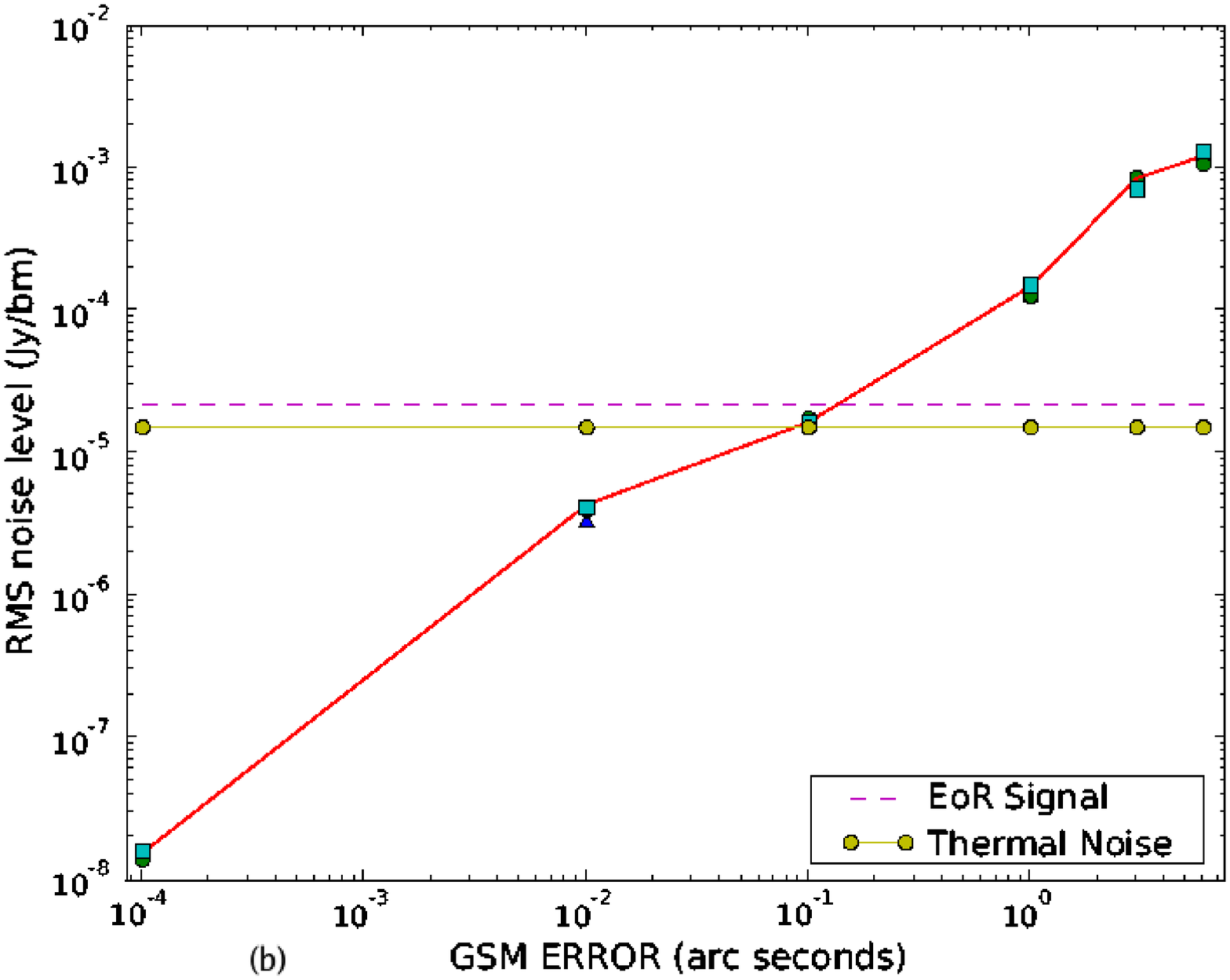,height=2.6truein}
\caption{{\bf (a)} The RMS noise variation in three different parts of
the images formed in the 3 steps of data reduction : [1] True Image
(Figure 4) $\rightarrow$ [2] UVSUB image (Figure 5a) $\rightarrow$ [3]
IMLIN image (Figure 5b). The GSM position error of 0.01 arc-seconds is
used here. {\bf (b)} The variation of RMS noise in the image plane with
the GSM error in position. Both the figures also include the
theoretical prediction for the EoR signal level ($22 ~\mu$~Jy/beam)
and the thermal noise ($15.4 ~\mu$~Jy/beam) after $5 \times 10^3$
hours of observations.}   
\end{figure*}

Flux errors might be as high as $\sim 1\%$, as quoted for most of the
source catalogs such as NVSS. The residuals will show up as sources
with flux density as high as few Jy/beam. In this case the residuals
are well above the level of the $S_{cut}$ and perfect removal of such
sources in the subsequent IMLIN step still remains a challenge.     

\subsection{Effect of Residual Calibration Errors - Simulations}
This section deals with the effect of residual calibration error
in the foreground source subtraction from the raw data-set coming out
of the upcoming EoR instruments.

The antenna dependent complex gains can be defined as:
\begin{equation}
V_{ij}^{Corr}(t) = g_i(t) g_j^*(t) V_{ij}^{o}(t)
\end{equation}
where $g_i$~'s are the antenna dependent complex gains, $V_{ij}^{Corr}$
are the corrupted visibilities and $V_{ij}^{o}$ are the true
visibilities or $V_{ij}^{GSM_{perfect}}$ in our case; $t$ denotes time
variation.  

The model for the complex gains used in the simulations is given by:
\begin{equation}
g_i(t)=(1+a_i+\delta a_i(t)) exp(i(\phi_i+\delta \phi_i(t))
\end{equation}
where $a_i$, $\phi_i$, $\delta a_i(t)$ and $\delta \phi_i(t)$ are
for amplitudes ($a$) and phases ($p$). The gains in both amplitude and
phase are derived from individual Gaussian random distribution as in
equation 6.

The process of simulation is as follows :

\begin{enumerate}
\item $V_{ij}^{Obs}$ are simulated. 

\item Model visibilities are predicted from a perfect GSM and are
corrupted with $g_i(t) \& g_j(t) $ to compute $V_{ij}^{Corr} $

\item In UVSUB we get : $V_{ij}^{Res}=V_{ij}^{GSM_{perfect}}(g_i g_j^*-1)$.

\item Make $I^{Res}$ and apply IMLIN as mentioned before.

\end{enumerate}

\begin{figure*}
\centering
\epsfig{file=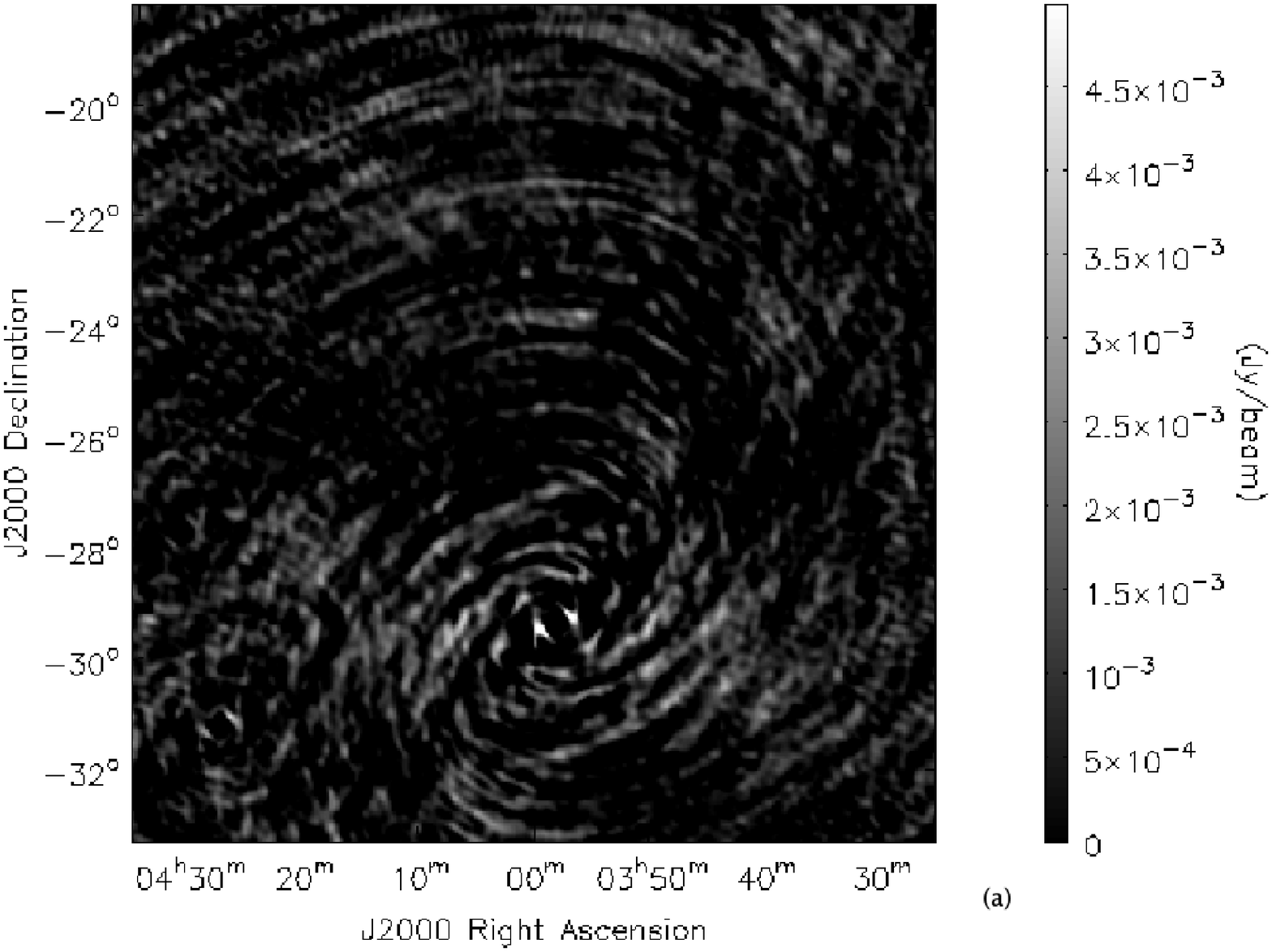,height=2.6truein}
\epsfig{file=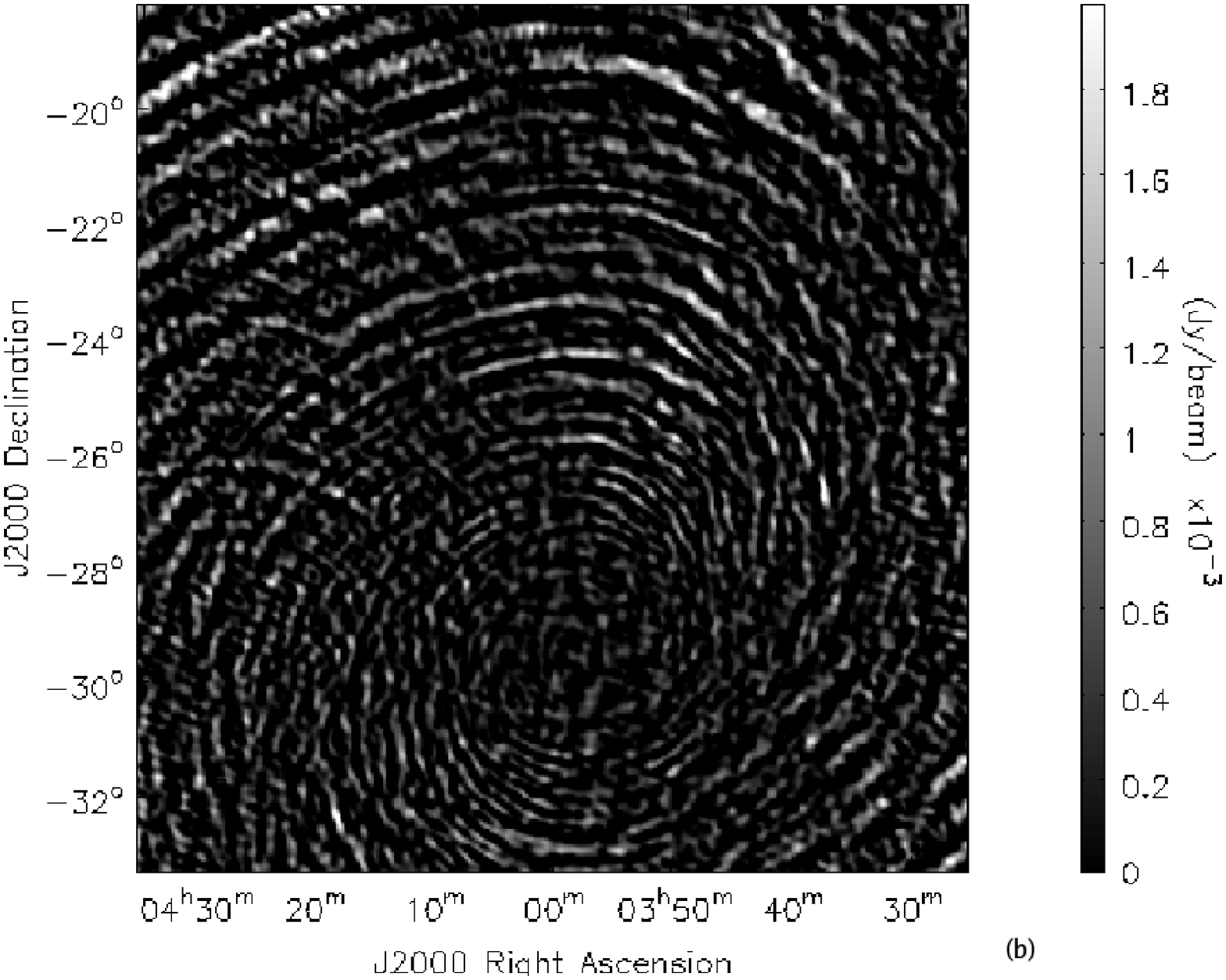,height=2.6truein}
\caption{{\bf (a)} Dirty UVSUBed image of the field made from the
  $V_{ij}^{res}$, after the model visibilities corrupted with residual
  calibration errors ($V_{ij}^{Corr}$) has been subtracted from the
  perfect data-set ($V_{ij}^{Obs}$). {\bf   (b)} The image of the
  field after the IMLIN has been applied to the UVSUBed image in
  figure a.}    
\end{figure*}


\begin{figure*}
\centering
\epsfig{file=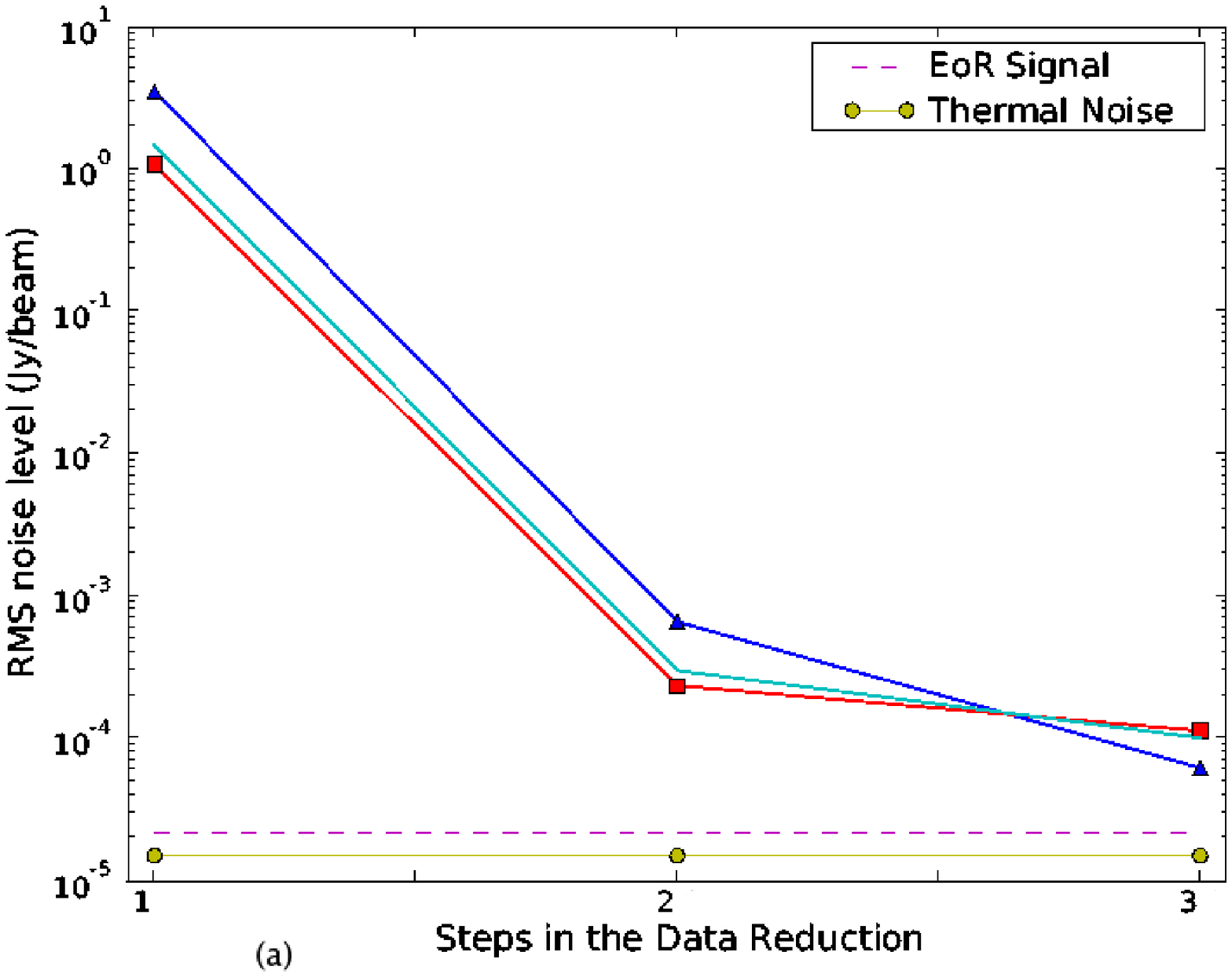,height=2.6truein}
\epsfig{file=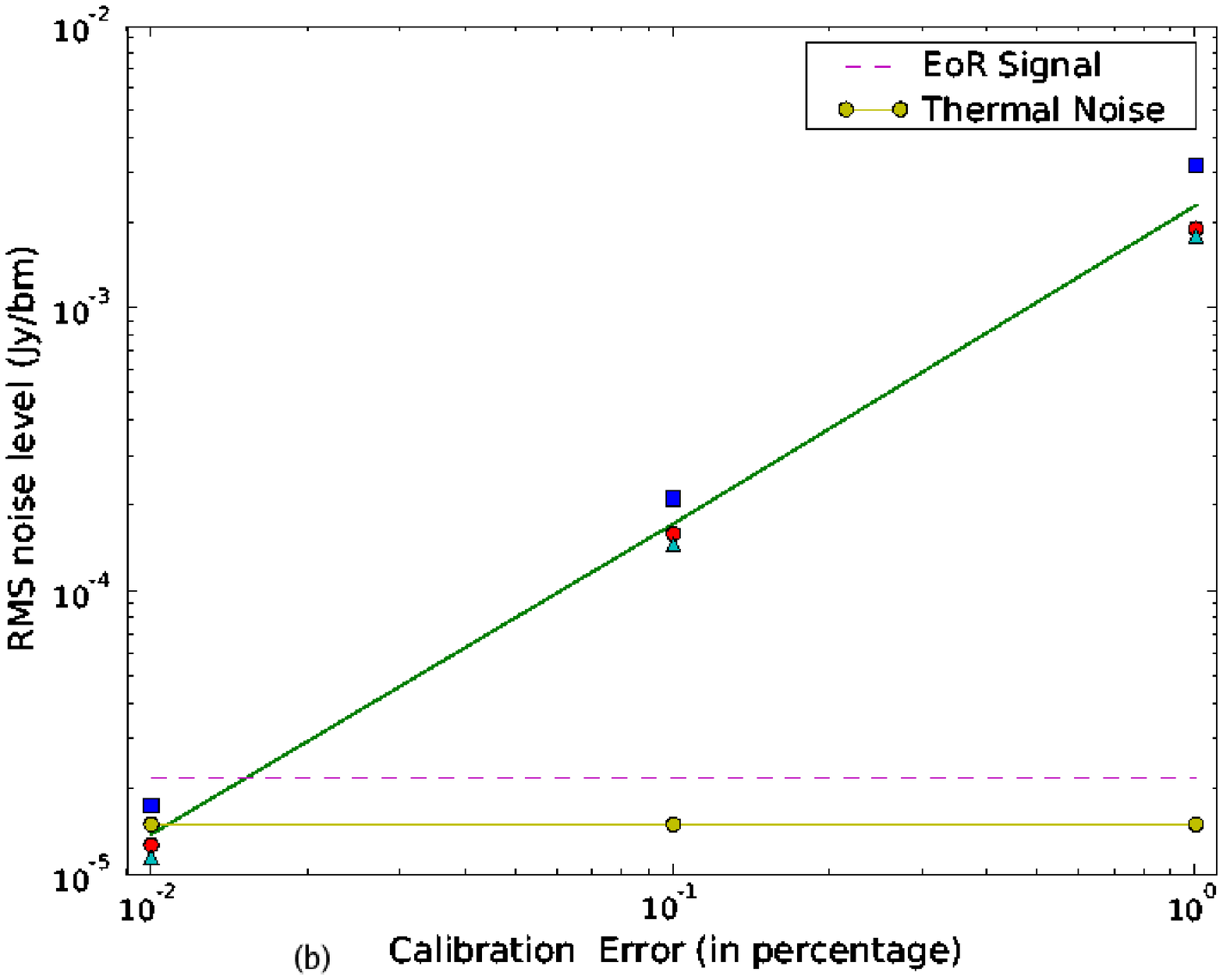,height=2.6truein}
\caption{{\bf (a)} The RMS noise variation in three different parts of
the images formed in the 3 steps of data reduction : [1] True Image
(Figure 4) $\rightarrow$ [2] UVSUB image (Figure 7a) $\rightarrow$ [3]
IMLIN image (Figure 7b). The calibration error used here is $0.1 \%$
in amplitude or 0.1 degree in phase. {\bf (b)} The variation of RMS
noise in the image plane with the Residual Calibration error. The
values denote the amplitude gain error in percentage or similar error
in degree for the phase gain. Both the figures also include the
theoretical prediction for the EoR signal level ($22 ~\mu$~Jy/beam) and
the thermal noise ($15.4 ~\mu$~Jy/beam) after $5 \times 10^3$ hours of
observations.} 
\end{figure*}

The model for the complex antenna-dependent gain errors (equation 7)
includes two terms for amplitude ($a_i$) and phase $\phi_i$ gains,
which are constant for each antenna throughout each day of observing
(6 hours). Along with them there are also two small additive random
offsets $\delta a_i ~\&~ \delta \phi_i$, which vary within a single
day of observation. Following \citet{perley99}, it has been determined
through several simulations that the constant offsets ($a_i, \phi_i$)
are the dominant terms compared to the time dependent terms. Moreover,
since the time variation for $\delta a_i ~\&~  \delta \phi_i$ are
derived from a Gaussian distribution similar to equation 6, the effect
of these terms decreases upon averaging over time. Hence, in our
simulations we have only included the constant gain errors $a_i~ \&~
\phi_i$ and redefined equation 8 as :   
\begin{equation}
g_i(t) \sim (1+a_i) e^{i \phi_i}
\end{equation}

It is certainly evident that some variant of the self calibration
algorithm will be implemented in the real-time calibration procedure
for the upcoming MWA like telescopes in order to facilitate the bright
foreground source removal. In our simulations, we have not implemented
any form of calibration algorithm. Instead, we have used the
measurement equations with residual calibration errors in order to
explore the accuracy with which any calibration procedure must work in
order to achieve the desired RMS noise limit and detect the signal from
reionization. 

\subsection{Effect of Residual Calibration Errors - Results}

Figure 7 demonstrates the three steps followed to simulate the
effect of the residual calibration errors in the EoR data-set. Figure
7a shows the dirty image of the residual UVSUB data and Figure 7b
shows the image after IMLIN. Both these figures are from the
simulation with $0.1\%$ amplitude error or $0.1$~degree phase error.

Figure 8a shows the variation in the RMS noise level in the three
regions of the images made using : [1] $V_{ij}^{Obs}$ (figure 4), [2]
$V_{ij}^{Res}$ in UVSUB (figure 7a) and [3] after IMLIN (figure
7b). It is clear from figure 8a that the RMS noise level reached near
the bright sources after the UVSUB stage is $\sim 10^{-3}$ Jy/beam,
corresponding to a dynamic range of $\sim 10^5$. \citet{perley99}
calculated the dynamic range limitation due to a time independent
amplitude error in all the baselines that will be introduced to a
point source data. If we denote the visibility function ($V_{ij}$) as
:  
\begin{equation}
V_{ij}(\vec{u})=(1+a)e^{i\phi}
\end{equation}
where $a ~\& ~\phi$ are the amplitude and phase errors respectively,
the dynamic range of the image is limited to :
\begin{equation}
DR \sim \frac{\sqrt{N(N-1)}}{\sqrt{2(a^2 +\phi^2)}} 
\end{equation}
where N is the number of antennas. In our case, $N=512$, $a=0.1\%$ and
$\phi=0.1$~degree. These values yield a dynamic range of $\sim 10^5$,
which agrees with the dynamic range we obtained from our simulations.

Figure 8b denotes the RMS noise value in different parts of the UVSUBed
image as a function of different residual calibration errors. The
three different residual calibration errors considered here are
$0.01\%$, $0.1\%$ and $1\%$ in amplitude or 0.01 degree, 0.1 degree
and 1 degree in phase respectively. The resultant curve in the log-scale
has a slope of $\sim 1.1$. This indicates that in order to achieve the
desired RMS noise level to detect the reionization signal, the residual
calibration errors should be $\sim 0.01\% $ in amplitude or $\sim
0.01$~degree in phase. In practice, it is extremely challenging to
achieve such accuracy with a real-time calibration procedure. Here, we
have only considered systematic residual calibration errors which will
not vary from one day to the other.   

As before, these simulations were originally performed for the
128-element array and then the RMS noise values were scaled down by a
factor $\sim 5$ following the scaling property that has been discussed
in initial sections. However, we have also repeated the simulations
for the 512-element array  with residual calibration error of $0.1 \%$
in amplitude or  0.1~degree in phase and found the results consistent
with our scaled values.   

\subsection{Reducing the effect of the residual calibration error}

In the previous section it is clearly shown that even the low
systematic residual calibration errors restrict the dynamic range in
the residual image to $10^5$, whereas the desired dynamic range is
$\sim 10^8$ in order to detect the signal from reionization. 

The obvious next step to follow is to add the UVSUBed image cubes from
successive days in order to check whether the residual calibration
error reduces as $\propto 1/\sqrt{N_{days}}$, where $N_{days}$ denotes
the number of days over which the UVSUB images has been added and each
day consists of 6 hours of observations. We can demonstrate the same
from the expression of the residual visibilities
($V_{ij}^{Res}=V_{ij}^{GSM_{perfect}}(g_i g_j^*-1)$). If the
calibration procedure is able to remove all systematic gain errors
from the corrupted data contributing to residual calibration errors
that are purely random from day to day, then they will reduce upon
averaging.      

\begin{figure*}[t!]
\centering
\epsfig{file=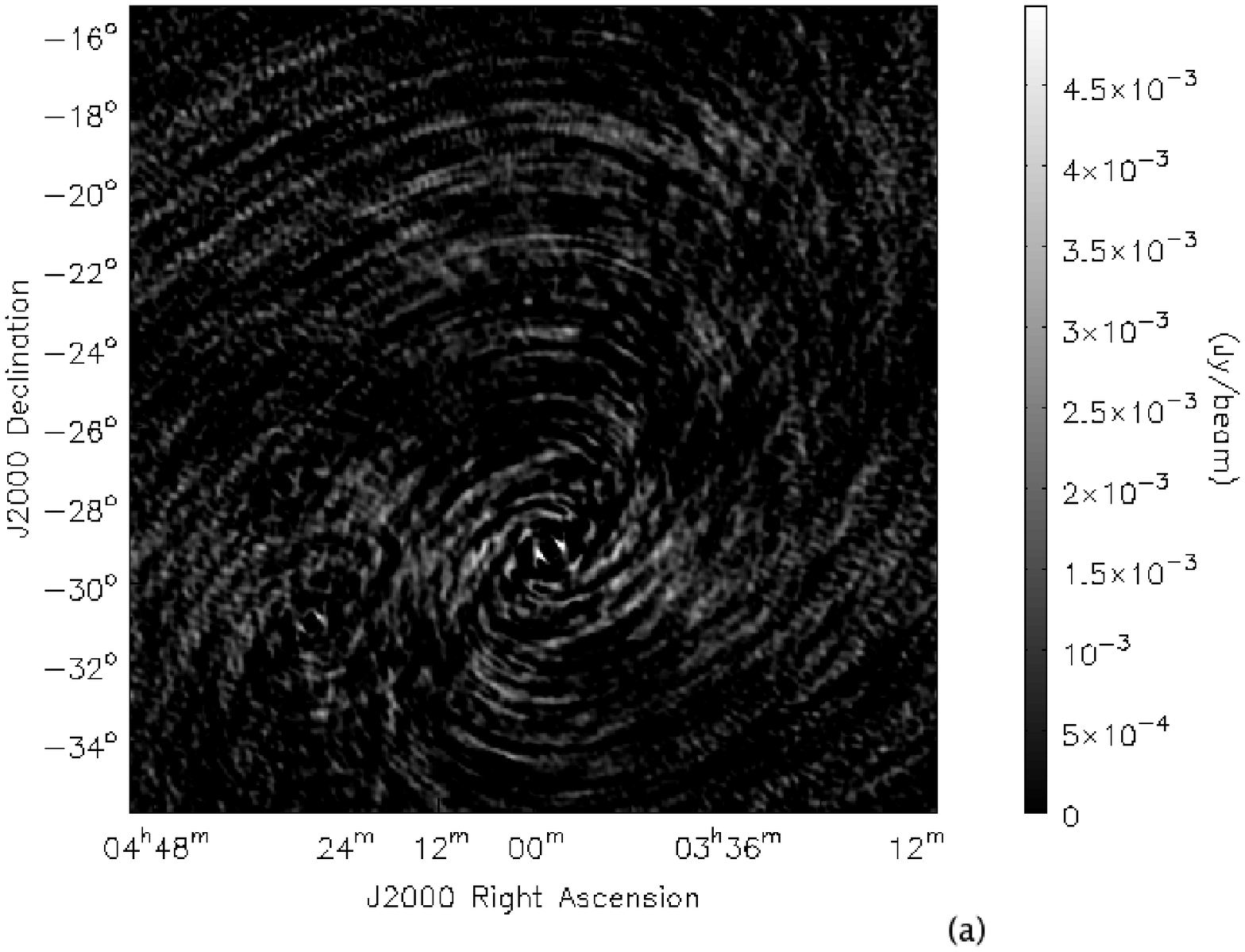,height=2.6truein}
\epsfig{file=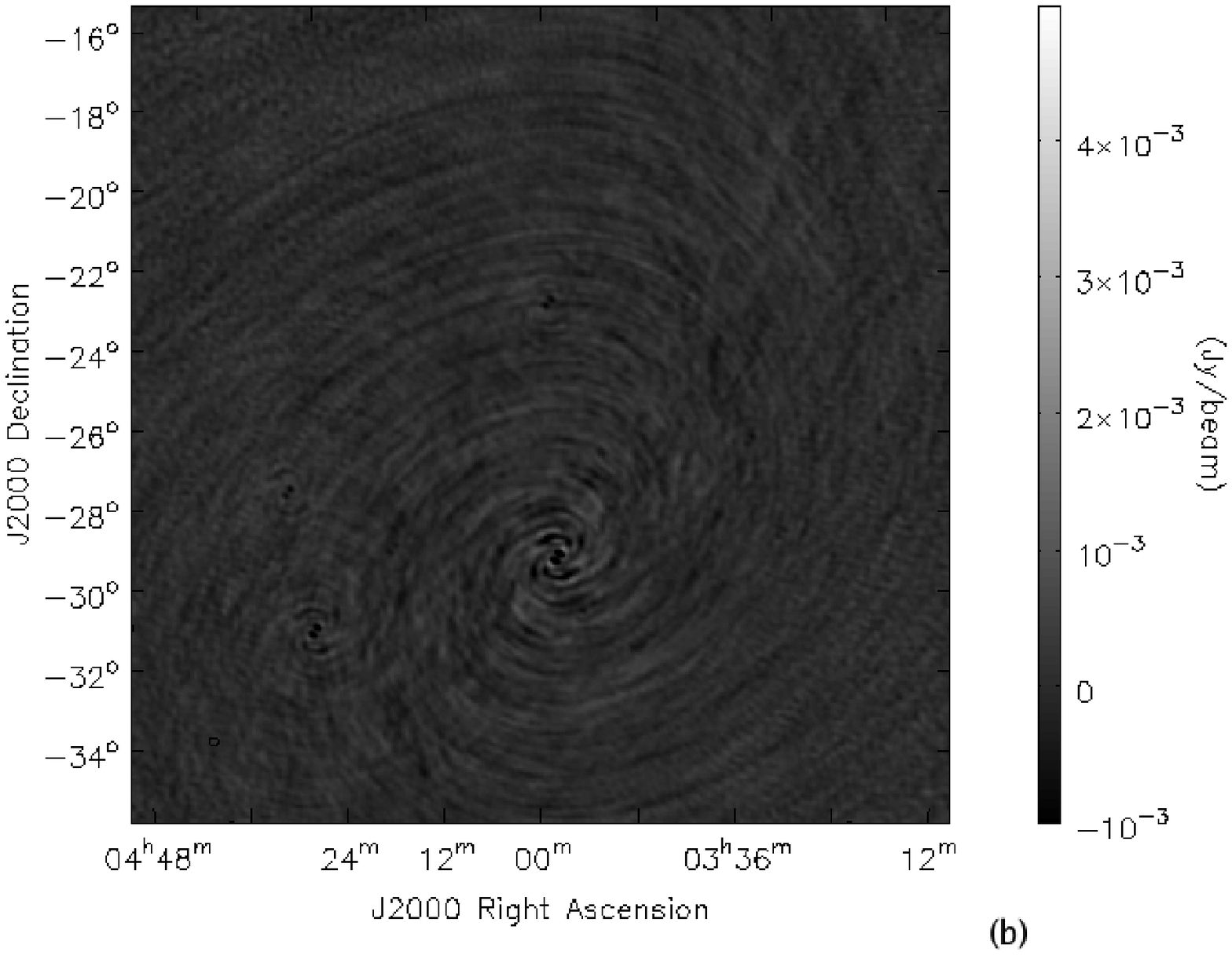,height=2.6 truein}
\caption{{\bf (a)} Dirty UVSUBed image of the field made from the
  $V_{ij}^{res}$, after the model visibilities corrupted with residual
  calibration errors ($V_{ij}^{Corr}$) has been subtracted from the
  perfect data-set ($V_{ij}^{Obs}$). {\bf   (b)} The averaged dirty
  image of the field after averaging 20 days of individual UVSUBed
  image, such as in figure a.}    
\end{figure*}

\begin{figure*}
\centering
\epsfig{file=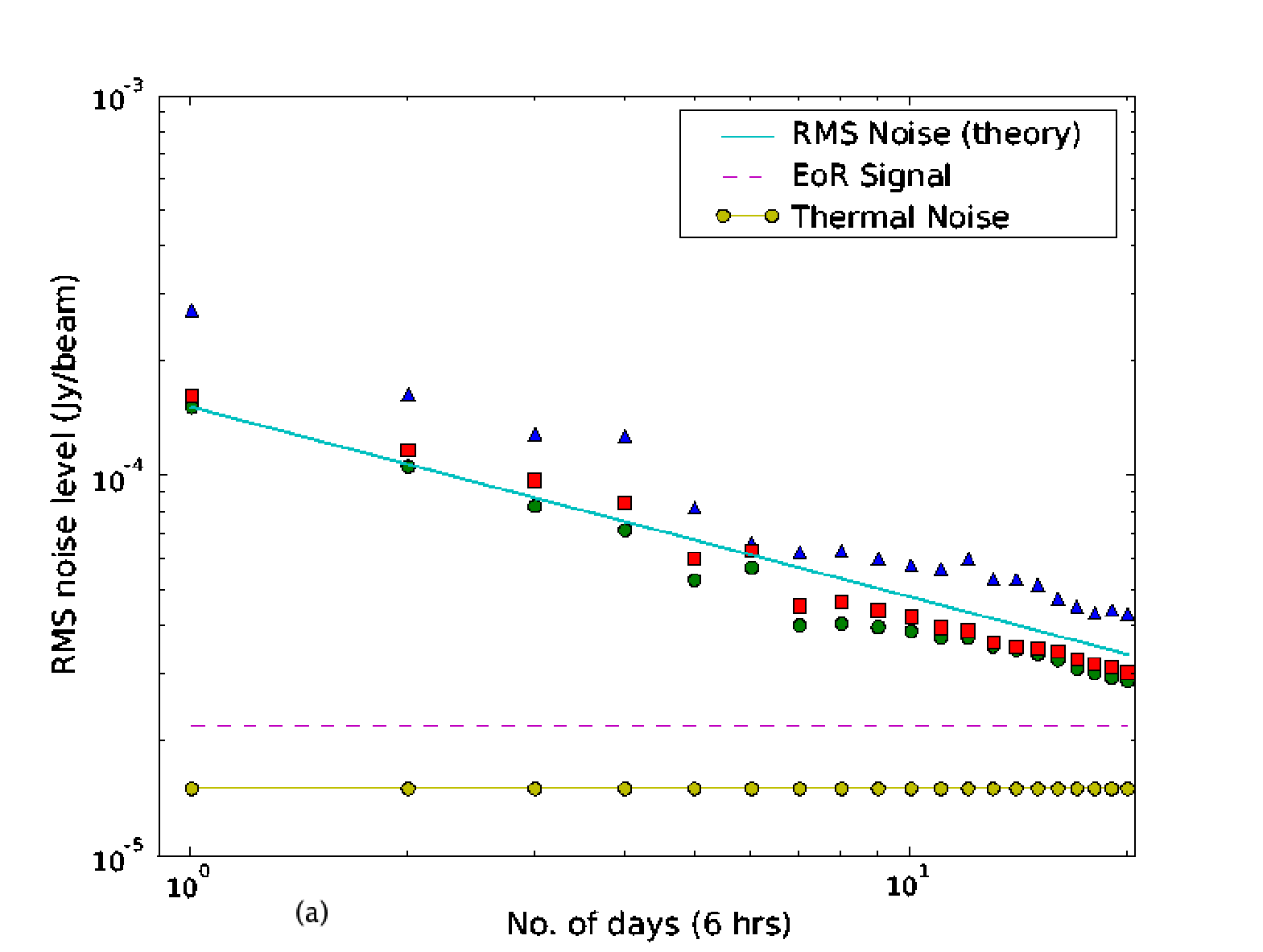,height=2.6truein}
\epsfig{file=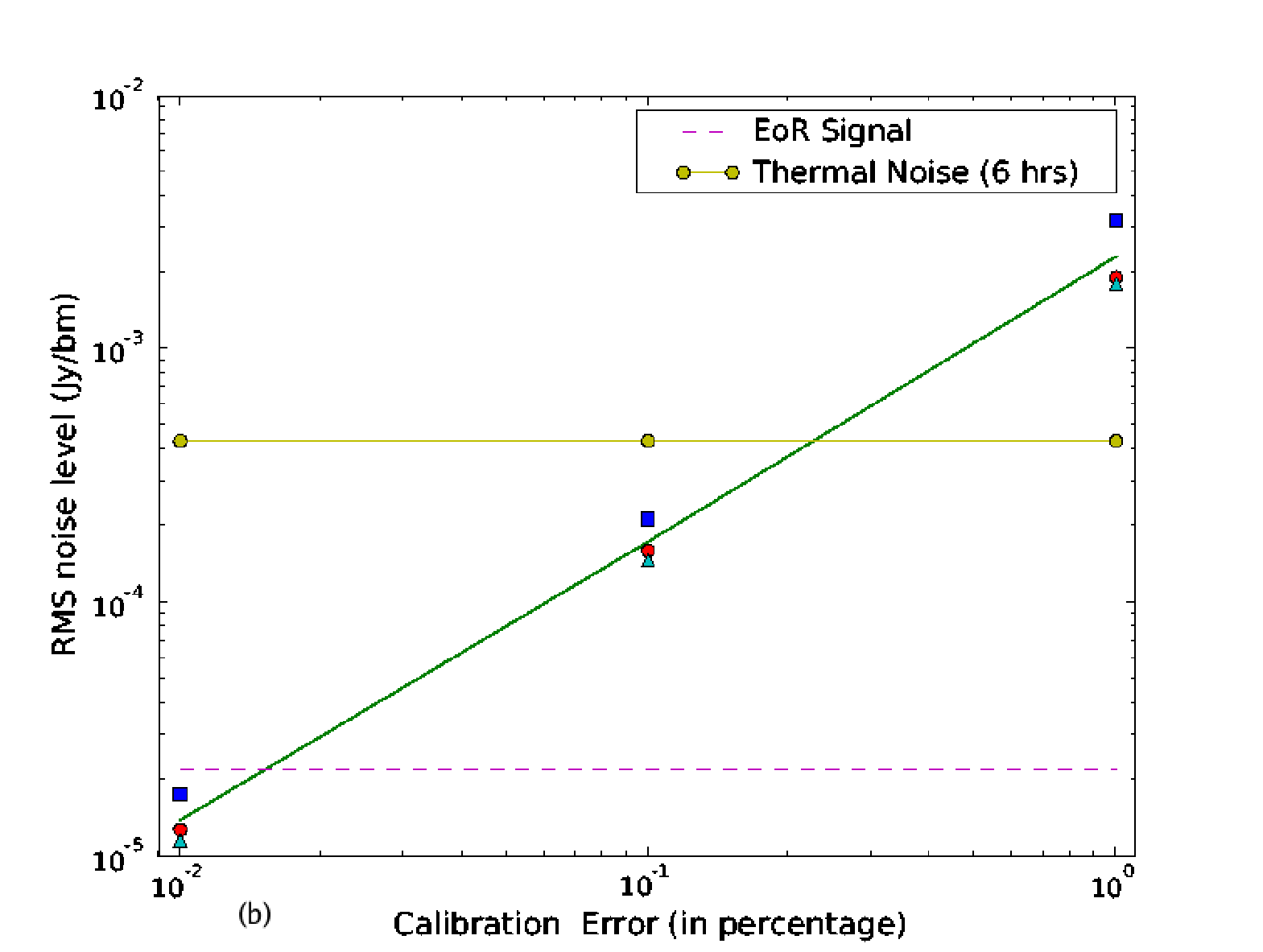,height=2.6truein}
\caption{{\bf (a)} The RMS noise variation in three different part of
the field shown as a function of the days of integration. The solid
line denotes the theoretical curve, showing a $\propto
1/\sqrt{N_{days}}$ dependence of the RMS noise reduction. The figure
also includes the theoretical prediction for the EoR signal level ($22
~\mu$~Jy/beam) and the thermal noise ($15.4 ~\mu$~Jy/beam) after $5
\times 10^3$ hours of observations. {\bf (b)} The variation of RMS
noise in the image plane with the Residual Calibration error. The
values denote the percentage error in the amplitude gain and similar
error in degree for the phase gain. The thermal noise quoted here is
$433 ~\mu$~Jy from 6 hours of integration (1 day).} 
\end{figure*}

Figure 9 demonstrates the effect of the averaging over 20 days of
UVSUBed images with the residual calibration errors of $0.1 \%$ in
amplitude or $0.1$~degree in phase. Figure 9a denotes the dirty image
after the first day and figure 9b denotes the averaged image after
20 days. The figures demonstrate a significant reduction in the
RMS noise level after 20 days of averaging. Figure 10a shows the
reduction in the RMS noise level in different parts of the image on
averaging over $20$ days. The solid line denotes the theoretical noise
curve that follows $\propto 1/\sqrt{N_{days}}$. Hence, the reduction
in RMS noise on averaging over different days is consistent with the
theoretical prediction ($\propto 1/\sqrt{N_{days}}$).

A UVSUBed image from each day (6 hours) is also limited by the
thermal noise, which reduces down as $\propto 1/\sqrt{N_{days}}$. If
the residual calibration errors are random between successive
data-sets from different days of integration then it should also
reduce down as $\propto 1/\sqrt{N_{days}}$. As a result, it should
then be sufficient to achieve a resultant RMS noise level in the
images below the daily thermal noise limit. From Table 1 we can
estimate that the thermal noise limit from 1 day of integration (6
hours) is $\sim 433~ \mu$~Jy/beam. Figure 9b is the same as figure 8b
except that we have plotted the new noise estimates from 6 hours of
observation. This figure now explains that the residual calibration
errors should be $ \sim 0.2\%$ in amplitude or $\sim 0.2$~degree in
phase in order to reach below the daily thermal noise limit. Once the
RMS noise level in the UVSUBed image is below the daily thermal noise
limit, it can reduce down along with the thermal noise in order to
detect the faint reionization signal within the time limit set by the
telescope sensitivity (i.e. $5 \times 10^3$~hours).    

The most important caveat in the entire analysis is the fact that we
assume that the residual calibration errors are not systematic from
one day to the next. If there are any systematics left over from day
to day (e.g. due to antenna gain errors, antenna primary beam errors,
pointing errors, etc) then those will not go down as $\propto
1/\sqrt{N_{days}}$. Moreover, even with these assumptions the much
needed accuracy in calibration is alarmingly high.

\section{Discussion - Implications on the upcoming arrays}

There are few important results and concerns that have come out from
the simulations performed in this paper.  

The simulations have been performed for both 128-element array and
512-element array. Necessary comparisons between both arrays have been
discussed at essential stages of this paper. The comparison between
the two configuration can be well explained by a simple scaling
relation which is the ratio between the RMS value of the PSF side-lobe
levels in two configurations. Hence it was sufficient to perform the
512-element simulations only for the optimal value of the GSM position
error and residual calibration error.  

The choice of $S_{cut} = 1$~Jy in our GSM is not totally 
arbitrary keeping in mind the fact that the real-time calibration
procedure needs to model and remove all the sources down to $S_{cut}$
from the raw UV data-sets within a duration constrained by the
instrumental specifications. Since the errors we have discussed in
this paper are dominated by the bright sources inside the
field-of-view, our analysis will not change significantly by reducing
the $S_{cut}$ level and including more and more fainter sources.  

Our simulations show two major results about the accuracy required to
reach the desired RMS noise level in order to detect the reionization
signal :

\begin{itemize}
\item GSM position accuracy of $\sim 0.1$~arc-second

\item Tolerance of residual calibration errors of $\lesssim 0.2 \%$ in
amplitude or $\lesssim 0.2$~degree in phase.
\end{itemize}

In the section dealing with the GSM position errors, we have only
discussed about systematic errors which do not change from one day
to the next. If the calibration procedure involves some iterative
position calibration scheme that will evaluate the source position
everyday then the errors from those procedure should vary from one day
to the other. Hence, there is a possibility of reaching the desired
RMS noise level even with lesser accuracy in the GSM position
estimates. 

In practice, a better GSM can be obtained a priori from a deep\
survey of the specific fields to be observed with MWA, LOFAR,
etc. using the present facilities like GMRT
\footnote{http://www.gmrt.ncra.tifr.res.in}  and WSRT
 \footnote{http://www.astron.nl} \citep{bernardi09} at 150 MHz. MWA,
 LOFAR, etc can also improve the existing GSM after sufficient days of
 observations.

The simulations with calibration errors assume that the image cubes
are formed after each day of observations (6 hours). We assume that
the calibration errors are incoherent beyond 6 hours and as a result
the effect of calibration error will reduce with successive days of
integration. If calibration error have shorter coherence time, then
lesser accuracy in the calibration can be tolerated in order to reach
below the thermal noise limit for the shorter integrations. However,
the systematics need to be removed at those timescales to make sure
that the residual errors are still random from one data-set to the
next. It should be noted that the PSF side-lobe level will also be
higher due to shorter integration period, which in turn might make it
harder to solve and remove the systematics at a shorter timescales.

The data analysis pipeline adapted in this paper resembles mostly the
upcoming MWA telescope. Since the raw visibilities cannot be stored,
the calibration and removal of the brightest sources in the real-time
calibration pipeline is one of the major step in foreground
subtraction \citep{mitchell08}. The accuracy in GSM position and
residual calibration (discussed in this paper) is critical for
extraction of the EoR signal. In other telescopes like LOFAR,
  where it is possible to store the raw visibilities, the calibration
  will be carried out through a combination of real-time processing
  and off-line reprocessing scheme. The ability to store the raw
  visibilities will allow more accurate calibration of the data in the
  off-line mode. However, the combination of real-time and
  off-line calibration pipeline for LOFAR still needs to achieve
  similar accuracy in calibration (discussed in this paper) in order
  to reach the desired dynamic range to extract the EoR signal.

In the present paper we have not discussed in great detail the step of
polynomial fitting in the image domain. This is mainly because the
lack of frequency dependence of the sky model as well as the
instrumental gain model. Our simulations do not include thermal
noise. In practice, the IMLIN step should be applied to the residual
image-cubes, averaged over total number of days required to beat
thermal noise. Hence, it should produce similar decrease in the RMS
noise level as obtained in this paper on ``noise-less'' data.  

The major feedback from all these simulations that we get is that the
raw UV-data should be retained, even considering the huge data-rate
issue, until the iterative real-time calibration procedures achieve
the desired accuracy. Upcoming telescopes like MWA, PAPER, LOFAR, etc
will also be trying to detect the EoR signal in the Power Spectral
domain. Hence, a relevant extension of our present work will be to
extend these analysis into the power spectral domain. We are also
planning to deal with frequency dependent and direction dependent
calibration errors in subsequent publications.

\acknowledgments 
\noindent AD and CC are grateful for support from the Max-Planck
Society and the Alexander von Humboldt Foundation through the Max
Planck Forshungspreise 2005. The National Radio Astronomy Observatory
is a facility of the National Science Foundation operated under
cooperative agreement by Associated Universities, Inc.

\bibliographystyle{apj}

\end{document}